\documentclass[a4paper,11pt,preprintnumbers]{article}
\pdfoutput=1 
\usepackage{jheppub} 

\usepackage[T1]{fontenc} 

\usepackage{amssymb}
\usepackage{amsmath}
\usepackage{bm}

\newcommand{\g}{\gamma}
\newcommand{\sig}{\sigma}

\newcommand{\sT}{\scriptstyle T}

\newcommand{\nn}{\nonumber}
\newcommand{\Pslash}{\kern 0.2 em P \kern -0.56 em \raisebox{0.3ex}{/}}
\newcommand{\Rslash}{\kern 0.2 em R \kern -0.56 em \raisebox{0.3ex}{/}}
\newcommand{\kslash}{\kern 0.2 em k \kern -0.45 em /}
\newcommand{\pslash}{\kern 0.2 em p \kern -0.45 em /}
\newcommand{\Sslash}{\kern 0.2 em S \kern -0.45 em /}
\newcommand{\nslash}{\kern 0.2 em n \kern -0.45 em /}
\newcommand{\thslash}{\kern 0.2 em \theta \kern -0.54 em /}

\newcommand{\p}{\perp}

\newcommand{\smarrow}{\mbox{\raisebox{-4.5pt}[0pt][0pt]{$\hspace{-1pt}\vec{\phantom{v}}$}}}
      
\newcommand{\hermes}{\textsc{Hermes }}

\newcommand{\belle}{\textsc{Belle }}

\newcommand{\bes}{\textsc{Bes-III }}

\newcommand{\bef}{\begin{figure}[htb]\centering}
\newcommand{\eef}{\end{figure}}
\newcommand{\beq}{\begin{equation}}
\newcommand{\eeq}{\end{equation}}

\newcommand{\as}{\alpha_s}
\newcommand{\half}{\textstyle{\frac{1}{2}}}

\newcommand{\avg}[1]{\big\langle  #1 \big\rangle}

\def\bea#1\eea{\begin{align}#1\end{align}}

\def\L{\Lambda}

\def\<{\langle}
\def\>{\rangle}

\def\a{\alpha}
\def\b{\beta}
\def\g{\gamma}  \def\G{\Gamma}
\def\d{\delta}  
   \def\L{\Lambda}

\def\m{\mu}
\def\mb{\mu_b}

\def\bmax{b_{\text{max}}}

\def\({\left(}
\def\[{\left[}
\def\){\right)}
\def\]{\right]}

\def\cos{\hbox{cos}}
\def\sin{\hbox{sin}}
\def\ln{\hbox{ln}}

\def\nslash{n\!\!\!\slash}

\def\pslash{p\!\!\!\slash}

\def\kslash{k\!\!\!\slash}

\def \le { \left    }
\def \ri { \right }

\def\lqcd{\L_{\rm QCD}}

\preprint{NIKHEF 2014-035}

\title{Effects of TMD evolution and partonic flavor on $e^+e^-$ annihilation into hadrons}

\author[a,b]{Alessandro Bacchetta,}
\author[c,d]{Miguel G. Echevarria,}
\author[c,d]{Piet J.G. Mulders,}
\author[a]{Marco Radici,}
\author[c,d]{Andrea Signori}

\affiliation[a]{INFN Sezione di Pavia,\\ via Bassi 6, I-27100 Pavia, Italy}
\affiliation[b]{Dipartimento di Fisica, Universita di Pavia, \\ via Bassi 6, I-27100 Pavia, Italy}
\affiliation[c]{Department of Physics and Astronomy, VU University Amsterdam,\\De Boelelaan 1081, NL-1081 HV Amsterdam, the Netherlands}
\affiliation[d]{Nikhef,\\Science Park 105, NL-1098 XG Amsterdam, the Netherlands}

\emailAdd{alessandro.bacchetta@unipv.it}
\emailAdd{miguelge@nikhef.nl}
\emailAdd{p.j.g.mulders@vu.nl}
\emailAdd{marco.radici@pv.infn.it}
\emailAdd{asignori@nikhef.nl}

\abstract{We calculate the transverse momentum dependence in the production of two back-to-back hadrons in electron-positron annihilations at the medium/large energy scales of \bes and \belle  experiments. We use the parameters of the transverse-momentum-dependent (TMD) fragmentation functions that were recently extracted from the semi-inclusive deep-inelastic-scattering multiplicities at low energy from \hermes. TMD evolution is applied according to different approaches and using different parameters for the nonperturbative part of the evolution kernel, thus exploring the sensitivity of our results to these different choices and to the flavor dependence of parton fragmentation functions. We discuss how experimental measurements could discriminate among the various scenarios.} 

\date{\today}

\begin{document}

\maketitle
\flushbottom
 
\section{Introduction}
\label{s:intro}

Transverse momentum dependent (TMD) parton distribution functions (PDFs) and fragmentation functions (FFs) depend on the longitudinal and transverse components of the momentum of partons with respect to the parent hadron momentum, as well as on their flavor and polarization state. The TMD PDFs and TMD FFs enlarge the amount of nonperturbative information carried by ordinary integrated PDFs and FFs because they open the window on explorations of the multi-dimensional structure of hadrons in momentum space in terms of their QCD elementary constituents. For example, in the last years several data for single- and double-spin asymmetries in semi-inclusive deep-inelastic scattering (SIDIS) have been accumulated and can be interpreted as originating from the effect of specific combinations of (polarized) TMD PDFs and TMD FFs (for a review, see Refs.~\cite{Bacchetta:2006tn,Barone:2010zz,Boer:2011fh,Aidala:2012mv}). 

The TMD PDFs and TMD FFs can be defined only by a careful selection of physical observables that are sensitive to processes with two separate scales. In addition one needs to study the appropriate factorization theorems for these observables. 
For example, the appropriate factorization theorem for SIDIS holds true if the hard photon virtuality is accompanied by transverse momenta of the order of nucleon mass~\cite{Ji:2004wu,Collins:2004nx},
which then are observed as a mismatch of collinear momenta. 
It is necessary that the definition of TMD functions includes all factorizable long-distance contributions to the physical cross section. These nonperturbative contributions, related to collinear gluon radiation, are summed into socalled gauge links that make the TMD functions color gauge invariant objects. Gauge links provide also the necessary phase to generate the above mentioned spin asymmetries~\cite{Brodsky:2002cx,Ji:2002aa,Belitsky:2002sm}. Because initial-state and final-state gluon interactions are summed into different gauge links, the TMD functions may be process dependent, although parity and time-reversal invariance can simplify this non-universality to a simple proportionality factor~\cite{Collins:2002kn,Boer:2003cm}. 
To account for scale dependence, the TMD functions obey evolution equations that generalize the standard Renormalization Group Evolution (RGE) to a multi-scale regime in hard processes. TMD evolution equations have been derived for unpolarized TMD PDFs and TMD FFs~\cite{Collins:2011zzd,Echevarria:2012pw}, and for polarized ones only in a limited number of cases~\cite{Aybat:2011ta,Bacchetta:2013pqa,Echevarria:2014rua}. But despite these recent achievements, the phenomenological implementation of these effects is still under active debate~\cite{Collins:2014loa,Echevarria:2014rua,Kang:2014zza,Angeles-Martinez:2015sea}. From the experimental point of view, only few data sets are available with enough statistics that allows for a multidimensional analysis and a direct access to transverse momentum distributions~\cite{Airapetian:2012ki,Adolph:2013stb}; in other cases, the studies were limited in the multidimensional coverage and by the restricted variety of targets and final-state hadrons~\cite{Arneodo:1984rm,Adloff:1996dy,Mkrtchyan:2007sr,Osipenko:2008aa,Asaturyan:2011mq}. 


In a preceding paper~\cite{Signori:2013mda}, 
the dependence of the intrinsic transverse-momentum distribution of both unpolarized TMD PDFs and TMD FFs upon the flavor and the longitudinal momentum of the parton involved was discussed using the recently published data from the \hermes 
collaboration~\cite{Airapetian:2012ki} on multiplicities for pions and kaons produced in SIDIS off proton and deuteron targets. 
Although the flavor-independent fit of the data was not statistically excluded, 
a clear indication was found 
that different quark flavors produce different transverse-momentum distributions of final hadrons, especially when comparing different species of final hadrons. This feature corresponds quite naturally to the well known strong flavor dependence of integrated 
PDFs~\cite{Forte:2013wc,Gao:2013xoa,Owens:2012bv,Ball:2012cx}, and to indications from some models~\cite{Bacchetta:2008af,Bacchetta:2010si,Wakamatsu:2009fn,Bourrely:2010ng,Matevosyan:2011vj,Schweitzer:2012hh} and lattice calculations of TMD objects~\cite{Musch:2010ka}. The SIDIS process is useful because it gives simultaneous access to TMD PDFs and TMD FFs. But the factorized cross section always involves
a convolution 
of transverse momenta of the initial and the fragmenting partons: anticorrelation hinders a separate investigation of the two intrinsic distributions. Moreover, the \hermes data were collected at such a limited range in the hard scale that the statistical analysis of 
Ref.~\cite{Signori:2013mda} was reasonably performed even without involving modifications due to evolution effects. 

In this paper, we consider the semi-inclusive production of two back-to-back hadrons in electron-positron annihilations. In analogy with the SIDIS process, we define the multiplicities in $e^+ e^-$ annihilations as the differential number of back-to-back pairs of hadrons produced per corresponding single-hadron production. Then, we study their transverse momentum distribution at large values of the center-of-mass (cm) energy, starting from an input expression for TMD FFs taken from the analysis of \hermes SIDIS multiplicities at low energy performed in Ref.~\cite{Signori:2013mda}. In this framework, we can extract clean and uncontaminated details on the transverse-momentum dependence of the unpolarized TMD FF, which is a fundamental ingredient of any spin asymmetry in SIDIS and, therefore, it affects the extraction also of polarized TMD distributions. Moreover, we can make realistic tests on the sensitivity to various implementations of TMD evolution available in the literature, since the hard scales involved in $e^+ e^-$ annihilations are much larger than the average values explored in SIDIS by \hermes , which is assumed as the starting reference scale. 

An important difference between PDFs and FFs is the role of the gauge links arising mostly from resummation of gluons with collinear polarizations. T-odd effects for PDFs enter through the operator definitions of the PDFs after inclusion of appropriate gauge links having also transverse pieces. For FFs T-odd effects are contained in the hadronic states and as a consequence there are less universality-breaking effects for FFs~\cite{Collins:2004nx,Gamberg:2008yt,Meissner:2008yf,Gamberg:2010uw}. 

In Sec.~\ref{s:theory}, we outline the theoretical tools needed to work out the cross sections for annihilations in two hadrons and define the $e^+ e^-$ multiplicities. In Sec.~\ref{s:evolution}, we introduce the QCD evolution of TMD FFs as the action of an evolution operator on input fragmentation functions, we describe some procedures to separate perturbative from nonperturbative domains of transverse momenta, and we provide some prescriptions to parametrize the nonperturbative contributions to the evolution kernel and the resummation of soft gluon radiation. In Sec.~\ref{s:flavor}, we introduce the flavor decomposition of fragmentation processes. In Sec.~\ref{s:results}, we make predictions for the spectrum in transverse momentum of $e^+ e^-$ multiplicities for production of two back-to-back hadrons, focusing on the sensitivity of results to the flavor of the fragmenting parton and to the different prescriptions for describing TMD evolution. Final comments and remarks are summarized in Sec.~\ref{s:end}.

\section{Multiplicities for $e^+ e^-$ annihilation into two hadrons}
\label{s:theory}

\bef
 \includegraphics[width=0.6\textwidth]{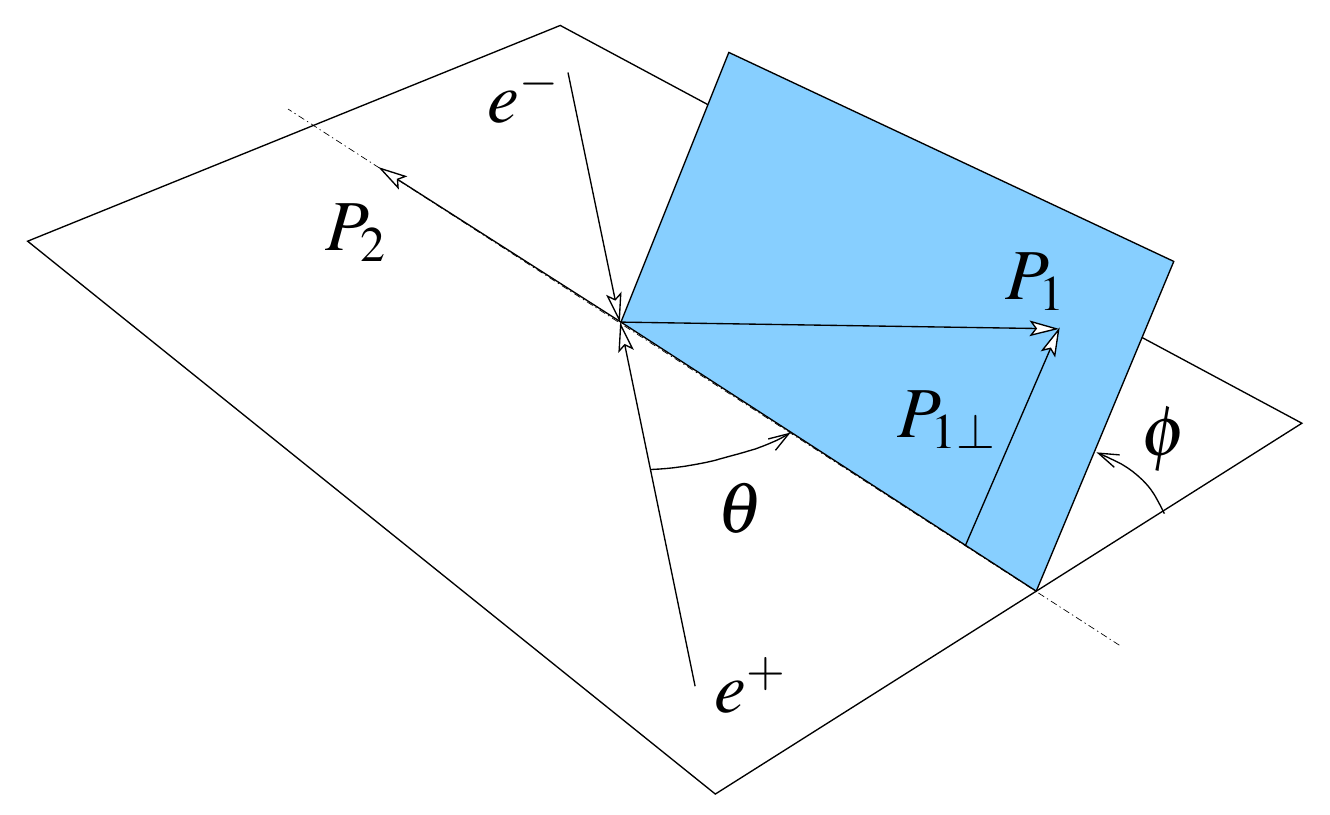}
  \caption{Kinematics for the $e^+ e^-$ annihilation leading to two back-to-back hadrons with momenta $P_1$ and $P_2$. }
  \label{f:kin}
\eef

We consider the process $e^+ e^- \to h_1 h_2 X$ depicted in Fig.~\ref{f:kin}. An electron $e^-$ and a positron $e^+$ annihilate producing a vector boson with time-like momentum transfer $q^2 \equiv Q^2 \geq 0$. A quark and an antiquark are then emitted, each one fragmenting into a residual jet containing a leading hadron that for simplicity we will consider unpolarized: the hadron $h_1$ with momentum and mass $P_1, M_1,$ and the hadron $h_2$ with momentum and mass $P_2,  M_2$. The two hadrons belong to two back-to-back jets, {\it i.e.} we have 
$P_1 \cdot P_2 \approx Q^2$. 
In the following, we will limit $Q^2$ values to a range where the vector boson can be safely identified with a virtual photon. Using the standard notations for the light-cone components of a 4-vector, we define the following invariants
\bea
z_1 =   \frac{2 P_1 \cdot q}{Q^2} \approx \frac{P_1^-}{q^-} \approx \frac{P_1 \cdot P_2}{q \cdot P_2}   \qquad  &
z_2= \frac{2 P_2 \cdot q}{Q^2} \approx \frac{P_2^+}{q^+} \approx \frac{P_2 \cdot P_1}{q \cdot P_1} \qquad  & 
y = \frac{P_2 \cdot \ell}{P_2 \cdot q} \; ,
\label{e:invariants}
\eea
where $\ell$ is the electron momentum. The $z_1$ is the fraction of parton momentum carried by the hadron $h_1$, and similarly for $z_2$ referred to the hadron $h_2$. Covariantly, we can define the normalized time-like and space-like directions
\beq
\hat t^\mu = \frac{q^\mu}{Q} \quad \mbox{and} \quad \hat z^\mu = \frac{Q}{P_2\cdot q}\, P_2^\mu - \hat t^\mu  =  \frac{2}{z_2 Q} P_2^\mu - \hat t^\mu \; . 
\eeq
Correspondingly, we can define  the projector into the space orthogonal to $\hat z$ and $\hat t$: 
\beq
g_\perp^{\mu\nu} = g^{\mu\nu} - \hat t^\mu\hat t^\nu + \hat z^\mu\hat z^\nu = g^{\mu\nu} - \frac{P_2^{\mu}q^{\nu} + q^{\mu}P_2^{\nu}}{P_2\cdot q} + O\left( \frac{M^2}{Q^2} \right) \; .
\eeq
The lepton momentum is then given by
\beq
\ell^\mu = \tfrac{1}{2} q^\mu + \left( y-\tfrac{1}{2}\right) Q \hat z^\mu + Q \sqrt{y(1-y)} \, \hat \ell_\perp^{\mu} \; ,
\eeq
where $\hat \ell_\perp^\mu = \ell_\perp^\mu / \vert \bm{\ell}_\perp \vert$ and $\ell_\perp^\mu = g_\perp^{\mu\nu} \ell_\nu$. 

The $g_\perp^{\mu\nu}$ projects onto the space orthogonal to $q$ and $P_2$. The projector onto the space orthogonal to $P_1$ and $P_2$, namely in the hadron cm frame where $P_1$ and $P_2$ have no transverse components, is given by
\bea
g_{\sT}^{\mu\nu} &= g^{\mu\nu} - \frac{P_1^{\mu}P_2^{\nu}+P_2^{\mu}P_1^{\nu}}{P_1\cdot P_2} + O\left( \frac{M^2}{Q^2} \right) \nn \\
&= g_\perp^{\mu\nu} + \frac{P_2^{\mu}q_{\sT}^{\nu}+q_{\sT}^{\mu}P_2^{\nu}}{P_2\cdot q} + O\left( \frac{M^2}{Q^2} \right) \; ,
\eea
where the non-collinearity is defined as
\bea
q_{\sT}^\mu &= q^\mu - \frac{P_1^\mu}{z_1} - \frac{P_2^\mu}{z_2} = g_{\sT}^{\mu\nu} q_{\nu} \nn \\
&= - \frac{P_{1\perp}^\mu}{z_1} + O\left( \frac{M^2}{Q^2} \right)  = - g_\perp^{\mu\nu} \frac{P_{1 \nu}}{z_1} + O\left( \frac{M^2}{Q^2} \right) \; .
\eea

In the electron-positron cm frame of Fig.~\ref{f:kin}, we define the angle $\theta = \arccos (\bm{\ell} \cdot \hat{\bm{z}} / |\bm{\ell}| )$ where 
$\hat{\bm{z}} = -\bm{P}_2$. It is related to the invariant $y \approx (1+\cos\, \theta)/2$. In analogy to the Trento 
conventions~\cite{Bacchetta:2004jz}, we define the azimuthal angle 
\beq
\cos\, \phi = 
        \frac{\bm{P}_2 \times \bm{\ell}}{|\bm{P}_2 \times \bm{\ell}|}
	\cdot 
	\frac{\bm{P}_{1\p}\times \bm{P}_2}{|\bm{P}_{1\p}\times \bm{P}_2|} 
 \; , 
\label{eq:az_angle}
\eeq
so that $P_1^{\mu} = \( 0, \, |\bm{P}_{1\p}|\, \cos\, \phi, \, |\bm{P}_{1\p}| \, \sin\, \phi , \, 0 \)$ in this frame, and in any frame obtained from this one by a boost along $\hat{\bm{z}}$. In general, the covariant definition is $\cos\,\phi = -q_{\sT}\cdot \hat \ell_\perp/\vert \bm{q}_{\sT}\vert$. 

The cross section for the $e^+ e^-$ annihilation into back-to-back pairs of unpolarized hadrons can be written in a factorized formula at low transverse momenta~\cite{Boer:1997mf,Collins:2011zzd,GarciaEchevarria:2011rb,Echevarria:2014rua}:
\bea
\frac{d\sig^{h_1 h_2}}{dz_1\, dz_2\, d q_{\sT}^2\, dy} &= \frac{6 \pi \a^2}{Q^2} \, A(y) \, {\cal H} (Q^2, \mu) \nn \\
&\hspace{-2cm} \times \sum_q \, e_q^2   \int_0^\infty d b_{\sT} \, b_{\sT} \, J_0 (q_{\sT} b_{\sT}) \, \[ z_1^2 \, D_1^{q\smarrow h_1} (z_1, b_{\sT}; \, \zeta_1, \, \mu) \, z_2^2 \, D_1^{\bar{q}\smarrow h_2} (z_2, b_{\sT}; \, \zeta_2, \, \mu) + (q \leftrightarrow \bar{q}) \]  \nn \\
&\hspace{-1.5cm} + Y (q_{\sT}^2 / Q^2) + {\cal O} (M^2 / Q^2)  \; , 
\label{e:fullxsect}
\eea
where $q_{\sT} \equiv |\bm{q}_{\sT}|$ and $A(y) = \half -y + y^2$. The ${\cal H}$ is the hard annihilation part. The 
$D_1^{q\smarrow h} (z, b_{\sT}; \, \zeta, \, \mu)$ is the TMD FF in impact parameter space for an unpolarized quark with flavor $q$ fragmenting into an unpolarized hadron $h$ and carrying light-cone momentum fraction $z$ and transverse momentum conjugated to $b_{\sT}$~\cite{Boer:2011xd}. Both ${\cal H}$ and $D_1^{q\smarrow h}$ are separated at the 
renormalization/factorization scale $\mu$ and evolve with it through renormalization group equations. The $D_1^{q\smarrow h}$ depends also on the scale $\zeta$ (with $\zeta_1 \zeta_2 = Q^4$) and evolves with it via a process-independent soft factor. The term $Y (q_{\sT}^2 / Q^2)$ ensures the matching with perturbative calculations at large transverse momenta. 

In this paper, we will consider a kinematics where $q_{\sT}^2 \ll Q^2$ and $M^2 \ll Q^2$. Hence, in Eq.~(\ref{e:fullxsect}) the 
$Y (q_{\sT}^2 / Q^2)$ term and corrections from higher twists of order $M^2 / Q^2$ or higher will be neglected. Moreover, the soft gluon radiation is here resummed into the TMD FF at the Next-to-Leading-Log level (NLL). It implies that the hard annihilation part is consistently calculated at leading order (LO) in $\as$, namely ${\cal H} (Q^2, \mu) \approx 1$. Equation~(\ref{e:fullxsect}) then simplifies to
\bea
\frac{d\sig^{h_1 h_2}}{dz_1\, dz_2\, d q_{\sT}^2\, dy} &\approx \frac{6 \pi \a^2}{Q^2} \, A(y)  \nn \\
&\hspace{-2cm} \times \sum_q \, e_q^2   \int_0^\infty d b_{\sT} \, b_{\sT} \, J_0 (q_{\sT} b_{\sT}) \, \[ z_1^2 \, D_1^{q\smarrow h_1} (z_1, b_{\sT}; \, \zeta_1, \, \mu) \, z_2^2 \, D_1^{\bar{q}\smarrow h_2} (z_2, b_{\sT}; \, \zeta_2, \, \mu) + (q \leftrightarrow \bar{q}) \]   \; . 
\label{e:xsect}
\eea

In Sec.~\ref{s:results}, we present our results for the $q_{\sT}$ spectrum of hadron pair multiplicities in $e^+ e^-$ annihilation. 
In strict analogy with the SIDIS definition~\cite{Airapetian:2012ki}, we construct the $e^+ e^-$ multiplicities as the differential number of back-to-back pairs of hadrons produced per corresponding single-hadron production after the $e^+ e^-$ annihilation. In terms of cross sections, we have 
\beq
M^{h_1 h_2} (z_1, z_2, q_{\sT}^2, y) = \frac{d\sig^{h_1 h_2}}{dz_1\, dz_2\, d q_{\sT}^2\, dy} \, {\Large /} \, 
\frac{d\sig^{h_1}}{dz_1\, dy} \; , 
\label{e:multi}
\eeq
where $d\sig^{h_1 h_2}$ is the differential cross section of Eq.~(\ref{e:xsect}). The $d\sig^{h_1}$ describes the production of a single hadron $h_1$ from the $e^+ e^-$ annihilation and it is obtained from the previous cross section by summing over all hadrons produced in one emisphere~\cite{Boer:1997mf}:
\beq
\frac{d\sig^{h_1}}{dz_1 dy} = \frac{12 \pi \a^2}{Q^2} \, A(y) \, \sum_q \, e_q^2  \, D_1^{q\smarrow h_1} (z_1) \; .
\label{e:1hxsect}
\eeq

\section{TMD evolution of fragmentation functions}
\label{s:evolution}


In the following, we describe in more detail the dependence of the fragmentation functions $D_1^{q\smarrow h}$ of Eq.~(\ref{e:xsect}) upon the renormalization/factorization scale $\mu$ and the scale $\zeta$. Different scenarios are possible according to the choice of the initial starting value for the factorization scale, and of the low-energy model describing the nonperturbative part of the evolution kernel. We first describe the structure of the input $D_1^{q\smarrow h}$ at the starting scale.

\subsection{Input fragmentation functions at the starting scale}
\label{ss:input}

We consider the unpolarized TMD FF extracted by fitting the hadron multiplicities in SIDIS data at low energy from \hermes 
~\cite{Airapetian:2012ki}. The assumed functional form displays a transverse-momentum dependent part which is described in impact parameter space by the following fixed-scale flavor-dependent Gaussian 
ansatz\footnote{The $1/z^2$ factors appearing in Eq.~(\ref{e:TMDFFQ0}) are due to $b_T$ being conjugated to the partonic transverse momentum $\bm{k}_{\sT}$, whereas the TMD FFs in Ref.~\cite{Signori:2013mda} are defined and normalized in momentum space with respect to the hadronic transverse momentum $\bm{K}_{\sT} = -z \bm{k}_{\sT}$.}:
\beq
D_1^{a\smarrow h} (z, b_{\sT}; \, Q^2) = d_1^{a\smarrow h} (z; \, Q^2) \, \frac{1}{z^2} \, \exp \[ - \frac{1}{4 z^2}\, 
\avg{\bm{P}^2_{\p}}^{a\smarrow h} (z) \, b_{\sT}^2 \]  \; ,
\label{e:TMDFFQ0}
\eeq
where $\avg{\bm{P}^2_{\p}}^{a\smarrow h} (z)$ with $a=q, \bar{q}$, is the flavor- and $z$-dependent Gaussian width at some starting scale $Q_0^2$~\cite{Signori:2013mda,Signori:2013gra,Signori:2014kda}. The choice of having separate Gaussian functions for different flavors is motivated by the significant differences displayed by the \hermes data between pion and kaon final-state hadrons~\cite{Airapetian:2012ki}. The factorized collinear dependent part $d_1^{a\smarrow h} (z; \, Q^2)$ is described by using the DSS parametrization of Ref.~\cite{deFlorian:2007aj}. 

Following Refs.~\cite{Nadolsky:2000ky,Landry:2002ix}, a possible energy dependence of the Gaussian distribution was taken into account introducing the logarithmic term 
\beq
\exp \bigg\{ -g_2\ \frac{b_{\sT}^2}{4}\ \ln \frac{Q^2}{Q_0^2}\bigg\}  \; , 
\label{e:GaussQ}
\eeq
with $g_2$ a free parameter. Choosing $Q_0^2 = 1$ GeV$^2$, it was soon realized that the best-fit value for $g_2$ was compatible with zero. As a matter of fact, the $Q^2$ range spanned by \hermes is small and the obtained experimental data for multiplicities are not sensitive to evolution effects. For this reason, the fit was performed by using Eq.~(\ref{e:TMDFFQ0}) at a scale fixed to the experimental average value, namely $Q^2 = Q_0^2 = 2.4$ GeV$^2$. With this choice, the possible energy dependence of Eq.~(\ref{e:GaussQ}) is automatically eliminated. 

In summary, the input to our studies on the evolution of $D_1^{a\smarrow h}$ with the scales $\mu$ and $\zeta$ is referred to the expression in Eq.~(\ref{e:TMDFFQ0}) to be considered at the starting scale $Q_0^2 = 2.4$ GeV$^2$. However, depending on the choice of the initial value of the factorization scale this identification is not always straightforward, as will be explained in the following sections. 

\subsection{The $\mb$ prescription}
\label{ss:collins_evo}

As shown in Eq.~(\ref{e:xsect}), the TMD FFs generally depend on the factorization scale $\mu$ and on the scale $\zeta$, that for convenience we name the rapidity scale. The TMD FFs satisfy evolution equations with respect to both of 
them~\cite{Collins:2011zzd,Echevarria:2012pw}. The evolution with respect to $\mu$ is determined by standard RGE equations, whereas the evolution in $\zeta$ is determined by a process-independent soft factor~\cite{Collins:2011zzd,Echevarria:2012pw}.

The functional form of TMD FFs at small $b_{\sT}$ can be calculated in perturbative QCD. Conversely, the nonperturbative part at large $b_{\sT}$ must be constrained by fitting experimental data. At the medium/large energies of the \bes and \belle experiments, the perturbative tail of TMD FFs needs to be taken into account. Using the technique of Operator Product Expansion (OPE), it can be represented as a convolution of (perturbatively calculable) Wilson coefficients $C$ with the (nonperturbative) collinear fragmentation functions $d_1$~\cite{Collins:2011zzd,Echevarria:2012pw}:
\beq
D^{a \smarrow h}(z, b_{\sT}; \zeta, \mu) = \underbrace{[C \otimes d^{a \smarrow h}_1](z, b_{\sT}; \zeta,\mu)}_{\text{small}\ b_{\sT}} + \underbrace{{\cal O}(b_{\sT} \Lambda_{\text{QCD}})}_{\text{large}\ b_{\sT}} \;  .
\label{e:ope_tmd}
\eeq

The convolution is defined as
\beq
[C \otimes d_1^{a \smarrow h}](z, b_{\sT}; \zeta,\mu) = \sum_{j=q,\bar{q}, g} \int_z^1 \frac{ds}{s}\ C_{j \smarrow a}\( \frac{z}{s}, b_{\sT};\zeta, \mu \) \  d_1^{j \smarrow h}(s; \mu) \; .
\label{e:convolution}
\eeq
The dependence of the coefficients upon both factorization and rapidity scales can be represented in a factorized form:
\beq
C_{j \smarrow a}(z, b_{\sT}; \zeta, \mu) = \bigg( \frac{\zeta}{\mb^2} \bigg)^{- K(b_{T}; \mu)}\  C_{j \smarrow a}(z, b_{\sT}; \mb^2, \mu) \; ,
\label{e:coefficient}
\eeq
where $\mb$ is defined as
\beq
\mb = \frac{2 e^{-\gamma_E}}{b_{\sT}} \; , 
\label{e:def_mub}
\eeq
and $\g_E$ is the Euler constant. 
The $K$ function in Eq.~(\ref{e:coefficient}) 
\footnote{Our $K$ function corresponds to the $D$ function in Ref.~\cite{Echevarria:2012pw}, and to the $\tilde{K}$ function in Ref.~\cite{Collins:2011zzd} but for a factor $-1/2$.}
arises from the process-independent soft factor that is necessary to proof the factorization theorem leading to the definition of the TMD FFs; it drives the evolution of TMD FFs in the $\zeta$ variable. 
The convolution in Eq.~(\ref{e:convolution}) is only valid for small $b_{\sT}$, namely $b_{\sT} \ll 1/\lqcd$. Moreover, the expression of the 
$C$ coefficients consists in a power series in $\as\  \ln\ (\mu^2 / \mb^2)$ (including also double logarithms of the same argument). The OPE is valid only when the logarithms do not diverge; this is accomplished, {\it e.g.}, by choosing $\mu = \mb$ or $q_{\sT}$, so that the series converges. Accordingly, if we choose $\mu = \mb$ we can write the TMD FF as
\bea
D^{a \smarrow h} (z, b_{\sT}; \zeta, \mb) &=  \( \frac{\zeta}{\mb^2} \)^{-K (b_{T}; \mb)}
\sum_{j=q,\bar{q}, g} \int_{z}^{1}\frac{ds}{s} \   C_{j \smarrow a}\( \frac{z}{s}, b_{\sT}; \mb^2, \mb \) \  d_1^{j \smarrow h} (s; \mb) \nn \\
&+ {\cal O} (b_{\sT} \lqcd ) \; .
\label{e:tmdff_ope}
\eea

The evolution of this fragmentation function from $\mb$ to another value of $\mu$ (e.g., $\mu = Q$) is driven by RGE equations. Instead, the evolution from an initial rapidity scale $\zeta_i$ to $\zeta$ is controlled by the $K$ function. The final expression of the TMD FF at the scales $\mu = Q$ and $\zeta$ is
\bea
D^{a \smarrow h}(z, b_{\sT}; \zeta, Q) &= \exp \le\{ \int_{\mb}^{\mu=Q} \frac{d\bar{\mu}}{\bar{\mu}} \gamma_{FF} \ri\} 
\( \frac{\zeta}{\zeta_{i}} \)^{-K(b_{T}; \mb)} \nn \\
&\times \( \frac{\zeta_i}{\mb^2} \)^{-K(b_{T}; \mb)}  \sum_{j=q,\bar{q}, g}  \int_{z}^{1} \frac{ds}{s} \  C_{j\smarrow a} \( \frac{z}{s}, b_{\sT}; \mb^2, \mb \) \  d_1^{j \smarrow h}(s; \mb) \nn \\
& + {\cal O} (b_{\sT} \lqcd ) \; ,
\label{e:evo_tmdff1}
\eea
where the anomalous dimension $\gamma_{FF}$ reads
\beq
\gamma_{FF} = -  \( \G_{\text{cusp}} \  \ln \frac{\zeta}{\mu^2} + \g^V \) \; ,
\label{e:gammaff}
\eeq
and $\G_{\text{cusp}}$ and $\g^V$ are also power series in $\as$ in the $\overline{\text{MS}}$ scheme~\cite{Echevarria:2014rua}. 

The above procedure is valid up to a maximum value of $b_{\sT}$, that we name $\bmax$, beyond which we do not trust the perturbative calculation. Hence, it is convenient to reconsider the OPE by introducing the new variable $\hat{b}_{\sT}$ that freezes at $\bmax$ when $b_{\sT}$ becomes large:
\beq
\lim_{b_{T} \to \infty} \hat{b}_{\sT} (b_{\sT}) = \bmax \; .
\label{e:bhat}
\eeq
For $b_{\sT} \lesssim \bmax$, the evolution in $\zeta$ is controlled by the function $K(\hat{b}_{\sT}; \mu_{\hat{b}})$, where 
\beq
\mu_{\hat{b}} = \frac{2 e^{-\gamma_E}}{\hat{b}_{\sT}} \; .
\label{e:def_mubhat}
\eeq
The nonperturbative part at large $b_{\sT}$ is defined as what is left over~\cite{Collins:2014jpa}:
\beq
g_{\text{np}}(b_{\sT}) = - K(\hat{b}_{\sT}; \mu_{\hat{b}}) + K(b_{\sT}; \mb) \; .
\label{e:def_np}
\eeq

By adding the intrinsic transverse distribution at the starting scale (see Eq.~(\ref{e:TMDFFQ0})), Eq.~(\ref{e:evo_tmdff1}) becomes
\bea
D^{a \smarrow h}(z, b_{\sT}; \zeta, Q) &= \exp \le\{ \int_{\mu_{\hat{b}}}^{Q} \frac{d\bar{\mu}}{\bar{\mu}} \gamma_{FF} \ri\}
\( \frac{\zeta}{\zeta_{i}} \)^{-K(\hat{b}_{T}; \mu_{\hat{b}}) - g_{\text{np}}(b_{T})}  \nn \\
&\hspace{-1cm} \times  \( \frac{\zeta_i}{\mu_{\hat{b}}^2} \)^{-K(\hat{b}_{T}; \mu_{\hat{b}}) - g_{\text{np}}(b_{T})}  
\sum_{j=q,\bar{q}, g} \int_{x}^{1} \frac{ds}{s}
C_{j \smarrow a} \( \frac{z}{s}, \hat{b}_{\sT}; \mu_{\hat{b}}^2, \mu_{\hat{b}} \) \  d_1^{a \smarrow h}(s; \mu_{\hat{b}}) \nn \\
&\hspace{-1cm} \times \frac{1}{z^2}\  e^{-\frac{\langle \bm{P}^2_{\p} \rangle^{a\smarrow h} (z)}{4 z^2} b_{T}^2} 
\( \frac{\zeta_i}{Q_0^2} \)^{-g_{\text{np}}(b_{T})} \;  .
\label{e:tmdff_evo_model}
\eea
If we insert $\zeta_i = \mu_{\hat{b}}^2$ and $\zeta = \mu^2 = Q^2$, the above equation reduces to
\bea
D^{a \smarrow h}(z, b_{\sT}; Q^2, Q) &= \exp \le\{ \int_{\mu_{\hat{b}}}^{Q} \frac{d\bar{\mu}}{\bar{\mu}} \gamma_{FF} \ri\}	
\( \frac{Q^2}{\mu_{\hat{b}}^2} \)^{-K (\hat{b}_{T}; \mu_{\hat{b}})-g_{\text{np}}(b_{T})} \nn \\
&\times  \ \sum_{j=q,\bar{q}, g} \int_{z}^{1} \frac{ds}{s}  \  C_{j \smarrow a} \( \frac{z}{s}, \hat{b}_{\sT}; \mu_{\hat{b}}^2, \mu_{\hat{b}} \) \  
d_1^{j \smarrow h} (s; \mu_{\hat{b}})   \nn \\
&\times \frac{1}{z^2} \  e^{ -\frac{\langle \bm{P}^2_{\p} \rangle^{a\smarrow h} (z)}{4 z^2} \  b_{T}^2 } 
\( \frac{\mu_{\hat{b}}^2}{Q_0^2} \)^{-g_{\text{np}}(b_{T})} \nn \\
&\equiv R(b_{\sT}; Q^2, Q, \mu_{\hat{b}}^2, \mu_{\hat{b}}) \  D^{a \smarrow h}(z, b_{\sT}; \mu_{\hat{b}}^2, \mu_{\hat{b}}) \; . 
\label{e:tmdff_evo_model_final}
\eea
Hence, the net effect of evolution can be represented as the action of an evolution operator $R$ on the input TMD FF evaluated at the  scale $\mu_{\hat{b}}$, which is running with $\hat{b}_{\sT}$. This peculiar feature grants that there is a smooth matching between the perturbative domain at small $b_{\sT}$ and the nonperturbative domain at large $b_{\sT}$. It is interesting to remark that from Eqs.~(\ref{e:tmdff_evo_model}) and (\ref{e:tmdff_evo_model_final}) we deduce that modelling the nonperturbative part affects the whole $b_{\sT}$ spectrum, not only the large $b_{\sT}$ region. 

In this paper, we resum the soft gluon radiation up to NLL contributions in $\ln\ (\mu/\mb)$, which corresponds to include terms linear in $\as$ in the perturbative expansion of $K$ and $\g^V$, and quadratic in the expansion of $\G_{\text{cusp}}$~\cite{Echevarria:2014rua}: 
\bea
K (b_{\sT}; \mu) &= \frac{C_F}{2\pi} \  \as \  \ln \frac{\mu^2}{\mb^2}  \; , \nn \\
\g^V &= - \frac{3 C_F}{2 \pi}\  \as \; , \nn \\
\G_{\text{cusp}} &= \frac{C_F}{\pi} \as \le\{ 1 + \frac{\as}{4\pi} \[ \( \frac{67}{9} - \frac{\pi^2}{3} \) C_A - \frac{20}{9} \  T_F \  n_f \] \ri\} \; , 
\label{e:andim}
\eea
where $C_A = N_c, \  C_F = (N_c^2 - 1)/ 2N_c,$ are the usual Casimir operators for the gluon and fermion representations of the color group SU$(N_c)$ with $N_c$ colors, and $T_F = n_f/2 $ with $n_f$ the number of active quark flavors. Consistently, the coefficients $C$ are computed at LO in $\as$, namely they reduce to $\delta$ functions such that Eq.~(\ref{e:tmdff_evo_model_final}) simplifies to
\bea
D^{a \smarrow h}(z, b_{\sT}; Q) &= \exp \le \{ \int_{\mu_{\hat{b}}}^{Q} \frac{d\bar{\mu}}{\bar{\mu}} \gamma_{FF\big{\vert}_{\text{NLL}}} \ri \} \( \frac{Q^2}{\mu_{\hat{b}}^2} \)^{-K_{\text{NLL}} (\hat{b}_{T}; \mu_{\hat{b}})-g_{\text{np}}(b_{T})}  \nn \\
&\times d_1^{a \smarrow h}(z; \mu_{\hat{b}}) \  \frac{1}{z^2}\ \exp \le\{-\frac{\avg{\bm{P}^2_{\p}}^{a\smarrow h} (z)}{4 z^2} b_{\sT}^2 \ri\} \( \frac{\mu_{\hat{b}}^2}{Q_0^2} \)^{-g_{\text{np}}(b_{T})} \;  .
\label{e:tmdff_evo_nll}
\eea

The definition of $g_{\text{np}}(b_{\sT})$ in Eq.~(\ref{e:def_np}) obviously implies that this function depends on $\bmax$, i.e. on the value of the impact parameter that sets the separation between the perturbative and nonperturbative regimes. Indeed, by perturbatively expanding $K(b_{\sT}; \mb)$ at lowest order we have~\cite{Collins:2014jpa}
\beq
g_{\text{np}}(b_{\sT}) \approx \frac{\alpha_s (\mu_{\hat{b}}) \, C_F}{\pi}\, \ln \left( 1 + \frac{b_{\sT}^2}{\bmax^2} \right) \; . 
\label{e:gnp1st}
\eeq
For $b_{\sT} \ll \bmax$, this expression recovers the quadratic parametrization $\textstyle{\frac{1}{2}} g_2 b_{\sT}^2$ adopted in the fits of Refs.~\cite{Landry:2002ix} and~\cite{Konychev:2005iy}, and it suggests that the parameter $g_2$ is not free but anticorrelated to $\bmax$, and proportional to $\bmax^2$ through a perturbatively calculable coefficient. The $g_{\text{np}}$ function accounts for the radiation of soft gluons emitted from a parton. A small (large) value of 
$\bmax$ implies that the QCD perturbative description is valid up to relatively small (large) $b_{\sT}$ values. Consequently, the amount of soft gluons emission is larger (smaller) and we expect a large (small) value for $g_2$.
More generally, this anticorrelation is motivated by the fact that both the exact function $K(b_{\sT}; \mb)$ and the TMD FF itself must not depend on the arbitrary choice of $\bmax$. So, $\bmax$ should not be regarded as a free parameter to be fitted to data, but it should be considered as an arbitrary scale that separates perturbative from nonperturbative regimes: changing $\bmax$ implies a rearrangement of all terms in Eq.~(\ref{e:tmdff_evo_model}) such that the TMD FF does not change~\cite{Collins:2014jpa}. 

For the purpose of this work, we will consider anticorrelated pairs of values for $\{ \bmax, g_2 \}$, inspired to the values adopted in Refs.~\cite{Landry:2002ix} and~\cite{Konychev:2005iy}. We will also explore different expressions for each one of the $\hat{b}_{\sT}$ and $g_{\text{np}}$ functions. For $\hat{b}_{\sT}$, our first choice is the socalled ``$b$-star'' prescription~\cite{Collins:2011zzd,Landry:2002ix}
\beq
\hat{b}_{\sT} \equiv b_{\sT}^* = \frac{b_{\sT}}{\sqrt{1+\frac{b_{T}^2}{\bmax^2}}} \; .
\label{e:bstar}
\eeq
The second choice is based on the exponential function
\beq
\hat{b}_{\sT} \equiv b_{\sT}^\dagger = \bmax \bigg\{ 1 - \exp \bigg[ - \frac{b_{T}^4}{\bmax^4} \bigg] \bigg\}^{\frac{1}{4}} \; ,
\label{e:bsun}
\eeq
that is steeper and it approaches the asymptotic constant $\bmax$ more quickly. 
For $g_{\text{np}}$, we choose a linear function of $b_T^2$ similarly to Refs.~\cite{Nadolsky:2000ky,Landry:2002ix,Konychev:2005iy} (see also Eq.~(\ref{e:GaussQ})):
\beq
g_{\text{np}}^{\text{lin}} (b_{\sT}) = \frac{g_2}{4}\ b_{\sT}^2 \; .
\label{e:lin_model}
\eeq
The second choice is suggested by Eq.~(\ref{e:gnp1st}):  
\beq
g_{\text{np}}^{\text{log}} (b_{\sT}) = g_2\ \ln\  \bigg( 1 + \frac{b_{\sT}^2}{4} \bigg) \; .
\label{e:log_model}
\eeq
This expression was considered also in Ref.~\cite{Aidala:2014hva}, and it reduces to Eq.~(\ref{e:lin_model}) for small $b_{\sT}$. 

In principle, we have four different combinations of prescriptions: $\{b_{\sT}^*,g_{\text{np}}^{\text{lin}}\}$, $\{b_{\sT}^*,g_{\text{np}}^{\text{log}}\}$, $\{b_{\sT}^\dagger,g_{\text{np}}^{\text{lin}}\}$, and $\{b_{\sT}^\dagger,g_{\text{np}}^{\text{log}}\}$. However, after some preliminary exploration we realized that some of them were producing redundant results. Therefore, they have been neglected. In summary, the transverse-momentum spectrum of the multiplicities in Eq.~(\ref{e:multi}) will be analyzed by varying the anticorrelated pair of parameters $\{ \bmax, g_2 \}$, and by considering only the two combinations $\{b_{\sT}^*,g_{\text{np}}^{\text{lin}}\}$ and $\{b_{\sT}^\dagger,g_{\text{np}}^{\text{log}}\}$. 

Finally, we remark that if we choose $Q = \mu_{\hat{b}}$ in Eq.~(\ref{e:tmdff_evo_nll}), {\it i.e.} if we switch off evolution effects, we should recover the Gaussian model expression of Eq.~(\ref{e:TMDFFQ0}) for the TMD FF at the initial scale $Q_0$. Formally, this is not the case because in the second line the collinear $d_1$ is evaluated at $\mu_{\hat{b}}$ and the term $(\mu_{\hat{b}}^2 /  Q_0^2 )^{-g_{\text{np}}(b_{T})}$ survives. However, the Gaussian model of Eq.~(\ref{e:TMDFFQ0}) is deduced by fitting the \hermes SIDIS data, whose kinematics overlaps the domain of very large $b_{\sT} \gg b_{\text{max}}$, namely where $\hat{b}_{\sT} \approx b_{\text{max}}$. If we use the prescription $\hat{b}_{\sT} \equiv b_{\sT}^*$ of Eq.~(\ref{e:bstar}), it is easy to check that for $b_{\text{max}} = 0.7$ GeV$^{-1}$  we have $\mu_{\hat{b}}^2 \approx Q_0^2 = 2.4$ GeV$^2$. Hence, the $D^{a \smarrow h}(z, b_{\sT}; \mu_{\hat{b}})$ of Eq.~(\ref{e:tmdff_evo_nll}) at $Q =  \mu_{\hat{b}}$ actually behaves like the $D^{a \smarrow h}(z, b_{\sT}; Q_0)$ of Eq.~(\ref{e:TMDFFQ0}) at the scale $Q_0$ and at very large $b_{\sT}$ values, or equivalently for very small parton transverse momenta. 

\subsection{The fixed-scale prescription}
\label{ss:fixedscale_evo}

In Eq.~(\ref{e:tmdff_evo_model_final}), we have expressed the evolved TMD FF at a scale $Q$ as the result of an evolution operator $R$ acting on the same TMD FF evaluated at the scale $\mu_{\hat{b}}$ running with $b_{\sT}$. Alternatively, we can fix the initial scale at the value $Q_i^2 = Q_0^2 = 2.4$ GeV$^2$ for the whole $b_{\sT}$ distribution: 
\beq
D^{a \smarrow h}(z, b_{\sT}; Q) = R(b_{\sT}; Q, Q_i) \  D^{a \smarrow h}(z, b_{\sT}; Q_i) \; . 
\label{e:eis_evo}
\eeq
With this choice, it is not possible to apply the OPE for calculating a perturbative tail to which the TMD FF should match at low $b_{\sT}$, as it was done in Eq.~(\ref{e:ope_tmd}): we need a model input over the whole $b_{\sT}$ spectrum. In our case,  
it is now very easy to identify the input TMD FF at the starting scale $Q_i$ with the Gaussian parametrization of Eq.~(\ref{e:TMDFFQ0}) at $Q_0$. Then, for $\mu_i^2 = \zeta_i = Q_i^2 = Q_0^2 = 2.4$ GeV$^2$ the TMD FF evolved at NLL up to a final scale $\mu^2 = \zeta \equiv Q^2$ becomes  
\bea
D^{a \smarrow h}(z, b_{\sT}; Q) &= \exp \le\{ \int_{Q_i}^{Q} \frac{d\bar\m}{\bar\m} \gamma_{FF}\big{\vert}_{\text{NLL}} \ri\} \  
\( \frac{Q^2}{Q_i^2} \)^{-K_{\text{NLL}} (b_{T}; Q_i)}  \nn\\
&\times d_1^{a \smarrow h}(z; Q_i) \  \frac{1}{z^2}\  \exp \left\{-\frac{\avg{\bm{P}^2_{\p}}^{a\smarrow h} (z)}{4 z^2} b_{\sT}^2 \ri\} \,  .
\label{e:tmdff_evo_Qi_nll}
\eea
The contribution from the $g_{\text{np}}$ term in the input distribution does not appear because of the choice of the starting scale $\zeta_i = Q_i^2 = Q_0^2$. 

The choice $\mu_i = Q_i$ of identifying the starting factorization scale with a fixed scale for the whole $b_{\sT}$ spectrum has important consequences also on the function $K$. From Eq.~(\ref{e:andim}), we can expand $K$ in powers of $\ln\ (\mu/\mb)$: if $\mu_i \neq \mb$, the series may not converge. One possible workaround is to apply the resummation technique to the $K$ function itself~\cite{Echevarria:2012pw}. Here, we will discuss two different prescriptions: computing $K$ from Ref.~\cite{Collins:2011zzd} at a fixed order in $\as$; or dressing $K$ by resumming large logarithms of the kind $\ln\ (\mu/\mb)$~\cite{Echevarria:2012pw}.  In the first case, $K$ is expanded in powers of $\as$; in the second case, the expansion is in $\as \  \ln\ (\mu/\mb)$. If $\mu_i = \mb$, the two expansions are the same. 

We will refer to the first choice as the "fixed-scale" prescription. Contrary to the prescription described in the previous section, there is no need to define an arbitrary scale $\bmax$ to separate perturbative from nonperturbative regimes. However, the function $K$ is evolved from $\mu_{\hat{b}}$ to $Q_i$ through its anomalous dimension:
\bea
K(b_{\sT}; Q_i) &= K(\hat{b}_{\sT}; \mu_{\hat{b}}) + \int_{\mu_{\hat{b}}}^{Q_i} \frac{d\bar{\mu}}{\bar{\mu}} \  \Gamma_{\text{cusp}} + g_{\text{np}}(b_{\sT}) \nn \\
&\stackrel{\approx}{\text{{\small NLL}}} \int_{\mu_{\hat{b}}}^{Q_i} \frac{d\bar{\mu}}{\bar{\mu}} \  \Gamma_{\text{cusp}} + g_{\text{np}}(b_{\sT})  \; ,
\label{e:K_collins}
\eea
where $g_{\text{np}}(b_{\sT})$ can get either the expression in Eq.~(\ref{e:lin_model}) or in Eq.~(\ref{e:log_model}). The perturbative contributions are calculated at NLL as in Eq.~(\ref{e:andim}), according to which we have $K_{\text{NLL}}(\hat{b}_{\sT}; \mu_{\hat{b}}) = 0$. 

The second choice is connected to the results of Ref.~\cite{Echevarria:2012pw}, because we resum all large logarithms of the kind $\ln\ (\mu/\mb)$ in the perturbative part as
\beq
K (b_{\sT}; Q_i) = D^R (b_{\sT}; Q_i) \  \theta (b_{\sT,c}-b_{\sT}) + \bar{g}_{\text{np}} (b_{\sT}) \  \theta (b_{\sT} - b_{\sT,c}) \; ,
\label{e:D_EISS}
\eeq
where $D^R$ is the resummed contribution computed in Ref.~\cite{Echevarria:2012pw}, and $b_{\sT,c}$ is the convergence radius of the perturbative expression. Apart from the resummation of logarithms, the main difference with Eq.~(\ref{e:K_collins}) is the presence of the $\theta$ functions: no $\hat{b}_{\sT}$ prescription is used to connect the perturbative and nonperturbative domains. And the nonperturbative contribution acts differently: while $g_{\text{np}}$ in Eq.~(\ref{e:K_collins}) applies to the  whole $b_{\sT}$ spectrum, in Eq.(\ref{e:D_EISS}) it does only for $b_{\sT} > b_{\sT,c}$. Hence, we use the notation $\bar{g}_{\text{np}}$ to account for this difference. For example, the $K$ function must be at least continuous at $b_{\sT} = b_{\sT,c}$. We can match this constraint by defining the nonperturbative contribution at $b_{\sT} > b_{\sT,c}$ as
\beq
\bar{g}_{\text{np}} (b_{\sT}) = D^R (b_{\sT,c})\ \bigg[ 1 + g_{\text{np}} (b_{\sT} - b_{\sT,c}) \bigg] \; ,
\label{e:gbar_def}
\eeq
where $g_{\text{np}}$ can be again either the $g^{\text{lin}}_{\text{np}}$ prescription of Eq.~(\ref{e:lin_model}) or the $g^{\text{log}}_{\text{np}}$ prescription of Eq.~(\ref{e:log_model}).

For $Q_i \ll Q$, the perturbative component $D^R$ in Eq.~(\ref{e:D_EISS}) diverges for $b_{\sT} < b_{\sT,c}$. Hence, its contribution to the evolution of the fragmentation function becomes negligible, being of the kind $\( Q / Q_i \)^{- D^R}$. Since $K$ is a smooth function in $b_{\sT}$, also the contribution of the nonperturbative part $\bar{g}_{\text{np}}$ for $b_{\sT} > b_{\sT,c}$ becomes numerically 
negligible~\cite{Echevarria:2012pw}. However, this result cannot be generalized to any  value of $Q$. Since we will make explorative calculations also at the \bes scale $Q = \sqrt{14.6}$ GeV which cannot be considered to be much larger than the initial scale $Q_0 = \sqrt{2.4}$ GeV of our input TMD FF, we will consider only the "fixed-scale" prescription of Eq.~(\ref{e:K_collins}). 

\subsection{Summary of evolution kernels}
\label{ss:evo_summary}

In summary, we consider two possible ways of evolving the TMD FF, according to the choice of the initial factorization scale $\mu_i$. It is understood that all formulae are computed at the NLL level of accuracy, according to Eq.~(\ref{e:andim}).

\begin{description}

\item[${\cal A}$ ] The "$\mb$" prescription: then $\mu_i^2 = \mu_{\hat{b}}^2 = \zeta_i, \; \zeta = \mu^2 = Q^2,$ and we have
\bea
D^{a \smarrow h} (z, b_{\sT}; Q) &= \exp \le\{ \int_{\mu_{\hat{b}}}^{Q} \frac{d\bar\m}{\bar\m} \gamma_{FF} \ri\} \  
\( \frac{Q^2}{\mu_{\hat{b}}^2} \)^{-K(\hat{b}_{T}; \mu_{\hat{b}}) - g_{\text{np}}(b_{T})}  \nn\\
&\times d_1^{a \smarrow h} (z; \mu_{\hat{b}}) \  \frac{1}{z^2}\  \exp \le\{-\frac{\avg{\bm{P}^2_{\p}}^{a\smarrow h} (z)}{4 z^2} b_{\sT}^2 \ri\} \  \( \frac{\mu_{\hat{b}}^2}{Q_0^2} \)^{-g_{\text{np}}(b_{T})} \; , 
\label{e:nCSS}
\eea
where $\g_{FF}$ and $K$ are described by Eqs.~(\ref{e:gammaff}) and~(\ref{e:andim}), $\mu_{\hat{b}}$ is given by Eq.~(\ref{e:def_mubhat}) with Eqs.~(\ref{e:bstar}) and~(\ref{e:bsun}), and $g_{\text{np}}$ is described in Eqs.~(\ref{e:lin_model}) and~(\ref{e:log_model}).

\item[${\cal B}$] The "fixed-scale" prescription: then $\mu_i^2 = Q_i^2 = \zeta_i, \, \zeta = \mu^2 = Q^2,$ and we have 
\bea
D^{a \smarrow h} (z, b_{\sT}; Q) &= \exp \le\{ \int_{Q_i}^{Q} \frac{d\bar\m}{\bar\m} \gamma_{FF} \ri\} \  
\( \frac{Q^2}{Q_i^2} \)^{-K (\hat{b}_{T}; \mu_{\hat{b}}) - \int_{\mu_{\hat{b}}}^{Q_i} \frac{d\bar\mu}{\bar\mu} \G_{\text{cusp}} - g_{\text{np}}(b_{T})}  \nn\\
&\times d_1^{a \smarrow h} (z; Q_i) \  \frac{1}{z^2}\ \exp \le\{-\frac{\avg{\bm{P}^2_{\p}}^{a\smarrow h} (z)}{4 z^2} b_{\sT}^2 \ri\} \; , 
\label{e:hybrid}
\eea
where $\g_{FF}, \, \mu_{\hat{b}}, \, g_{\text{np}}$ are defined in the same equations as above, while $K$ is given in Eq.~(\ref{e:K_collins}). 

\end{description}

\section{Flavor dependence of fragmentation functions}
\label{s:flavor}

The flavor sum in Eq.~(\ref{e:xsect}) can be made explicit and further simplified using the symmetry upon charge-conjugation transformations:  
\beq
D_1^{q \smarrow h} (z, b_{\sT}; \, Q^2) = D_1^{\bar{q} \smarrow \bar{h}} (z, b_{\sT}; \, Q^2) \; . 
\label{e:C-inv}
\eeq
At the starting scale $Q_0$, we distinguish the {\it favored} fragmentation where the fragmenting parton is in the valence content of the final hadron $h$. All the other channels are classified as {\it unfavored} fragmentation and are characterized by the fact that the detected hadron is produced by exciting more than one $q \bar{q}$ pair from the vacuum. If the final hadron is a kaon, we further distinguish a favored fragmentation initiated by an up quark/antiquark from the one initiated by a strange quark/antiquark. We limit the sum to three flavors $u, \, d, \, s$, and the corresponding antiquark partners. 

\subsection{Favored and unfavored fragmentation to different hadron species}
\label{s:fav-unfav}

For the final hadron pair being $(h_1, \, h_2) = (\pi^+, \, \pi^-)$, the flavor sum in Eq.~(\ref{e:xsect}) becomes
\beq
\sum_q \, e_q^2 \, D_1^{q\smarrow \pi^+} \, D_1^{\bar{q} \smarrow \pi^-} + (q \leftrightarrow \bar{q}) = D_{\rm fav}^{\pi^+ \pi^-} + 
D_{\rm unf}^{\pi^+ \pi^-}  \; , 
\label{e:flavpi+pi-}
\eeq
where
\bea
D_{\rm fav}^{\pi^+ \pi^-} (z_1, z_2, b_{\sT}; Q_0^2) &= \frac{4}{9} D_1^{u\smarrow \pi^+} (z_1, b_{\sT}; Q_0^2) \, D_1^{\bar{u}\smarrow \pi^-} 
(z_2, b_{\sT}; Q_0^2) \nn \\
&\quad + \frac{1}{9} D_1^{\bar{d}\smarrow \pi^+} (z_1, b_{\sT}; Q_0^2)\, D_1^{d\smarrow \pi^-} (z_2, b_{\sT}; Q_0^2) \; , 
\label{e:favpi+pi-}
\eea
and
\bea
D_{\rm unf}^{\pi^+ \pi^-} (z_1, z_2, b_{\sT}; Q_0^2) &= \frac{4}{9} D_1^{\bar{u}\smarrow \pi^+} (z_1, b_{\sT}; Q_0^2) \, D_1^{u\smarrow \pi^-} 
(z_2, b_{\sT}; Q_0^2) \nn \\
&\quad + \frac{1}{9} D_1^{d\smarrow \pi^+} (z_1, b_{\sT}; Q_0^2) \, D_1^{\bar{d}\smarrow \pi^-} (z_2, b_{\sT}; Q_0^2) \nn \\
&\qquad + \frac{1}{9} \( D_1^{s\smarrow \pi^+} (z_1, b_{\sT}; Q_0^2) \, D_1^{\bar{s}\smarrow \pi^-} (z_2, b_{\sT}; Q_0^2) \right. \nn \\
&\left. \qquad \qquad + D_1^{\bar{s}\smarrow \pi^+} (z_1, b_{\sT}; Q_0^2) \, D_1^{s\smarrow \pi^-} (z_2, b_{\sT}; Q_0^2) \) \; .
\label{e:unfpi+pi-}
\eea
Using the charge--conjugation symmetry of Eq.~(\ref{e:C-inv}), it is simple to prove that the result for 
$(h_1, \, h_2) = (\pi^-, \, \pi^+)$ is identical to the above one in Eq.~(\ref{e:flavpi+pi-}). 
 
If the final pions have the same charge, $(h_1, \, h_2) = (\pi^+, \, \pi^+)$, we have
\bea
\sum_q \, e_q^2 \, D_1^{q\smarrow \pi^+} \, D_1^{\bar{q} \smarrow \pi^+} + (q \leftrightarrow \bar{q}) &=  \[ \frac{4}{9} \, D_1^{u\smarrow \pi^+} 
(z_1, b_{\sT}; Q_0^2) \, D_{\rm unf}^{\pi^+} (z_2, b_{\sT}; Q_0^2) \right. \nn \\
&\left. \hspace{-1cm} + \frac{1}{9} \, D_{\rm unf}^{\pi^+} (z_1, b_{\sT}; Q_0^2) \, 
D_1^{\bar{d}\smarrow \pi^+} (z_2, b_{\sT}; Q_0^2) \] + (1 \leftrightarrow 2) \nn \\
&\; \; + \frac{2}{9} \, D_{\rm unf}^{\pi^+} (z_1, b_{\sT}; Q_0^2) \, D_{\rm unf}^{\pi^+} (z_2, b_{\sT}; Q_0^2) \; , 
\label{e:flavpi+pi+}
\eea
where
\bea
D_{\rm unf}^{\pi^+} (z, b_{\sT}; Q_0^2) &= D_1^{\bar{u}\smarrow \pi^+} (z, b_{\sT}; Q_0^2) = D_1^{d\smarrow \pi^+} (z, b_{\sT}; Q_0^2) \nn \\
&= D_1^{s\smarrow \pi^+} (z, b_{\sT}; Q_0^2) = D_1^{\bar{s}\smarrow \pi^+} (z, b_{\sT}; Q_0^2) \; .
\label{e:unfpi+}
\eea
Again, because of charge-conjugation symmetry we get the same result for $(h_1, \, h_2) = (\pi^-, \, \pi^-)$. 

If the final hadron pair is $(h_1, \, h_2) = (K^+, \, K^-)$, the flavor sum becomes
\beq
\sum_q \, e_q^2 \, D_1^{q\smarrow K^+} \, D_1^{\bar{q} \smarrow K^-} + (q \leftrightarrow \bar{q}) = D_{\rm fav}^{K^+ K^-} + 
D_{\rm unf}^{K^+ K^-}  \; , 
\label{e:flavK+K-}
\eeq
where
\bea
D_{\rm fav}^{K^+ K^-} (z_1, z_2, b_{\sT}; Q_0^2) &= \frac{4}{9} D_1^{u\smarrow K^+} (z_1, b_{\sT}; Q_0^2) \, D_1^{\bar{u}\smarrow K^-} 
(z_2, b_{\sT}; Q_0^2) \nn \\
&\quad + \frac{1}{9} D_1^{\bar{s}\smarrow K^+} (z_1, b_{\sT}; Q_0^2)\, D_1^{s\smarrow K^-} (z_2, b_{\sT}; Q_0^2) \; , 
\label{e:favK+K-}
\eea
and
\bea
D_{\rm unf}^{K^+ K^-} (z_1, z_2, b_{\sT}; Q_0^2) &= \frac{4}{9} D_1^{\bar{u}\smarrow K^+} (z_1, b_{\sT}; Q_0^2) \, D_1^{u\smarrow K^-} 
(z_2, b_{\sT}; Q_0^2) \nn \\
&\quad + \frac{1}{9} D_1^{s\smarrow K^+} (z_1, b_{\sT}; Q_0^2) \, D_1^{\bar{s}\smarrow K^-} (z_2, b_{\sT}; Q_0^2) \nn \\
&\qquad + \frac{1}{9} \( D_1^{d\smarrow K^+} (z_1, b_{\sT}; Q_0^2) \, D_1^{\bar{d}\smarrow K^-} (z_2, b_{\sT}; Q_0^2) \right. \nn \\
&\left. \qquad \qquad + D_1^{\bar{d}\smarrow K^+} (z_1, b_{\sT}; Q_0^2) \, D_1^{d\smarrow K^-} (z_2, b_{\sT}; Q_0^2) \) \; .
\label{e:unfK+K-}
\eea
Charge-conjugation symmetry grants the same result for $(h_1, \, h_2) = (K^-, \, K^+)$. 

If $(h_1, \, h_2) = (K^+, \, K^+)$:
\bea
\sum_q \, e_q^2 \, D_1^{q\smarrow K^+} \, D_1^{\bar{q} \smarrow K^+} + (q \leftrightarrow \bar{q}) &=  \[ \frac{4}{9} \, D_1^{u\smarrow K^+} 
(z_1, b_{\sT}; Q_0^2) \, D_{\rm unf}^{K^+} (z_2, b_{\sT}; Q_0^2) \right. \nn \\
&\left. \hspace{-1cm} + \frac{1}{9} \, D_{\rm unf}^{K^+} (z_1, b_{\sT}; Q_0^2) \, D_1^{\bar{s}\smarrow K^+} (z_2, b_{\sT}; Q_0^2) \] 
+ (1 \leftrightarrow 2) \nn \\
&+ \frac{2}{9} \, D_{\rm unf}^{K^+} (z_1, b_{\sT}; Q_0^2) \, D_{\rm unf}^{K^+} (z_2, b_{\sT}; Q_0^2) \; , 
\label{e:flavK+K+}
\eea
where
\bea
D_{\rm unf}^{K^+} (z, b_{\sT}; Q_0^2) &= D_1^{\bar{u}\smarrow K^+} (z, b_{\sT}; Q_0^2) = D_1^{s\smarrow K^+} (z, b_{\sT}; Q_0^2) \nn \\
&= D_1^{d\smarrow K^+} (z, b_{\sT}; Q_0^2) = D_1^{\bar{d}\smarrow K^+} (z, b_{\sT}; Q_0^2) \; .
\label{e:unfK+}
\eea
As before, we get the same result for $(h_1, \, h_2) = (K^-, \, K^-)$.

The last combination is $(h_1, \, h_2) = (\pi^+, \, K^-)$:
\bea
\sum_q \, e_q^2 \, D_1^{q\smarrow \pi^+} \, D_1^{\bar{q} \smarrow K^-} + (q \leftrightarrow \bar{q}) &=  D_{\rm fav}^{\pi^+ K^-} (z_1, z_2, b_{\sT}; Q_0^2) \nn \\
&\hspace{-4cm} + \frac{1}{9} \, \( D_1^{\bar{d}\smarrow \pi^+} (z_1, b_{\sT}; Q_0^2) \, D_{\rm unf}^{K^+} (z_2, b_{\sT}; Q_0^2) + 
D_{\rm unf}^{\pi^+} (z_1, b_{\sT}; Q_0^2) \, D_1^{s\smarrow K^-} (z_2, b_{\sT}; Q_0^2) \) \nn \\
&\hspace{-3.5cm} + \frac{2}{3} \, D_{\rm unf}^{\pi^+} (z_1, b_{\sT}; Q_0^2) \, D_{\rm unf}^{K^+} (z_2, b_{\sT}; Q_0^2) \; , 
\label{e:flavpi+K-}
\eea
where
\beq
D_{\rm fav}^{\pi^+ K^-} (z_1, z_2, b_{\sT}; Q_0^2) = \frac{4}{9} \, D_1^{u\smarrow \pi^+} (z_1, b_{\sT}; Q_0^2) \, D_1^{\bar{u}\smarrow K^-} 
(z_2, b_{\sT}; Q_0^2) \; , 
\label{e:favpi+K-}
\eeq
and charge-conjugation symmetry applied to Eq.~(\ref{e:unfK+}) gives 
\bea
D_{\rm unf}^{K^+} (z, b_{\sT}; Q_0^2) &= D_1^{u\smarrow K^-} (z, b_{\sT}; Q_0^2) = D_1^{\bar{s}\smarrow K^-} (z, b_{\sT}; Q_0^2) \nn \\
&= D_1^{\bar{d}\smarrow K^-} (z, b_{\sT}; Q_0^2) = D_1^{d\smarrow K^-} (z, b_{\sT}; Q_0^2) \; ,
\label{e:unfK-}
\eea
and grants that the same result in Eq.~(\ref{e:flavpi+K-}) holds also for $(h_1, \, h_2) = (\pi^-, \, K^+)$. 

\subsection{Flavor dependent Gaussian ansatz}
\label{s:gauss}

The starting input to our analysis are the TMD FFs extracted by fitting the hadron multiplicities in SIDIS data from \hermes at $Q_0^2 = 2.4$ GeV$^2$~\cite{Signori:2013mda}. The assumed functional form displays a transverse-momentum dependent part which is described in impact parameter space by the following flavor-dependent Gaussian ansatz:
\bea
D_1^{q\smarrow h} (z, b_{\sT}; \, Q_0^2) &= d_1^{q\smarrow h} (z; \, Q_0^2) \, \frac{1}{z^2} \, \exp \[ - \frac{1}{4 z^2}\, 
\avg{\bm{P}^2_{\p}}^{q\smarrow h} (z) \, b_{\sT}^2 \] \nn \\
&\equiv d_1^{q\smarrow h} (z; \, Q_0^2) \, G_q^h (z, b_{\sT}^2) \; .
\label{e:Gauss-ans}
\eea
The cross section of Eq.~(\ref{e:xsect}) (and, in turn, the multiplicity in Eq.~(\ref{e:multi})) is then a sum of Gaussians, and thus no longer a simple Gaussian. The width of the Gaussian depends also on the fractional momentum $z$, as done in several model calculations or phenomenological extractions~\cite{Boglione:1999pz,Bacchetta:2002tk,Schweitzer:2003yr,Bacchetta:2007wc,Matevosyan:2011vj,D'Alesio:2004up}. The chosen functional form is~\cite{Signori:2013mda}
\beq
\avg{\bm{P}^2_{\p}}^{q\smarrow h} (z) = \avg{\hat{\bm{P}}^2_{\p}}^{q\smarrow h} \, 
\frac{(z^\b + \d) \, (1-z)^{\g}}{(\hat{z}^\b + \d) \, (1-\hat{z})^{\g}} \; , 
\label{e:width}
\eeq
where $\b, \, \d, \, \g$, are fitting parameters and $\avg{\hat{\bm{P}}^2_{\p}}^{q\smarrow h} \equiv \avg{\bm{P}^2_{\p}}^{q\smarrow h} (\hat{z})$, with $\hat{z} = 0.5$. 

Isospin and charge-conjugation symmetries suggest four different Gaussian shapes~\cite{Signori:2013mda}:
\begin{gather}
\avg{\hat{\bm{P}}^2_{\p}}^{u \smarrow \pi^+} = \avg{\hat{\bm{P}}^2_{\p}}^{\bar{d} \smarrow \pi^+} = 
\avg{\hat{\bm{P}}^2_{\p}}^{\bar{u} \smarrow \pi^-} =  \avg{\hat{\bm{P}}^2_{\p}}^{d \smarrow \pi^-} \equiv 
\avg{\hat{\bm{P}}^2_{\p}}^{{\rm fav}} \; ,  
\label{e:favwidth}  
\\
\avg{\hat{\bm{P}}^2_{\p}}^{u \smarrow K^+} =  \avg{\hat{\bm{P}}^2_{\p}}^{\bar{u} \smarrow K^-} \equiv 
\avg{\hat{\bm{P}}^2_{\p}}^{uK}  \; ,
\label{e:uKwidth}  
\\
\avg{\hat{\bm{P}}^2_{\p}}^{\bar{s} \smarrow K^+} = \avg{\hat{\bm{P}}^2_{\p}}^{s \smarrow K^-} \equiv 
\avg{\hat{\bm{P}}^2_{\p}}^{sK} \;  ,
\label{e:sKwidth}  
\\
\avg{\hat{\bm{P}}^2_{\p}}^{\text{all others}} \equiv \avg{\hat{\bm{P}}^2_{\p}}^{{\rm unf}} \; .
\label{e:unfwidth} 
\end{gather} 
Correspondingly, we have four different Gaussian functions in Eq.~(\ref{e:Gauss-ans}): 
\bea
G_u^{\pi^+} = G_{\bar{d}}^{\pi^+} = G_{\bar{u}}^{\pi^-} = G_d^{\pi^-} &\equiv G_{\rm fav} (z, b_{\sT}^2)  \; , 
\label{e:favGauss}
\\
G_u^{K^+} = G_{\bar{u}}^{K^-} &\equiv G_{uK} (z, b_{\sT}^2) \; , 
\label{e:uKGauss}
\\
G_{\bar{s}}^{K^+} = G_s^{K^-} &\equiv G_{sK} (z, b_{\sT}^2) \; , 
\label{e:sKGauss}
\\
G_u^{\pi^-} =  G_u^{K^-} = G_d^{\pi^+} = G_d^{K^\pm} = G_s^{\pi^\pm} = G_s^{K^+} = G_{\bar{u}}^{\pi^+} = G_{\bar{u}}^{K^+} & \nn \\
= G_{\bar{d}}^{\pi^-} = G_{\bar{d}}^{K^\pm} = G_{\bar{s}}^{\pi^\pm} = G_{\bar{s}}^{K^-} &\equiv G_{\rm unf} (z, b_{\sT}^2) \; . 
\label{e:unfGauss}
\eea
Each one of these four functions depends on the same $\b, \, \d, \, \g$, fitting parameters of Eq.~(\ref{e:width}), such that all the $G_q^h (z, b_{\sT}^2)$ in Eq.~(\ref{e:Gauss-ans}) are described by seven parameters. 

For the collinear functions $d_1^{q\smarrow h} (z; \, Q_0^2)$, we adopt the same assumptions of Ref.~\cite{deFlorian:2007aj}:
\begin{itemize}
\item[-] isospin symmetry of the sea quarks

\item[-] for $h = \pi^+$, a direct proportionality between the $(d+\bar{d})$ and $(u+\bar{u})$ combinations, {\it i.e.} $(d+\bar{d}) = N (u+\bar{u})$. 
\end{itemize}
Therefore, we have three independent favored fragmentations, 
\begin{gather}
d_1^{u\smarrow \pi^+} (z; Q_0^2), \quad d_1^{u\smarrow K^+} (z; Q_0^2), \quad d_1^{\bar{s}\smarrow K^+} (z; Q_0^2)  \; , 
\label{e:favcoll}
\end{gather}
and two independent unfavored fragmentations: 
\begin{gather}
d_1^{\bar{u}\smarrow \pi^+} = d_1^{d\smarrow \pi^+} = d_1^{s\smarrow \pi^+} = d_1^{\bar{s}\smarrow \pi^+} \equiv d_{1\, {\rm unf}}^{\pi^+} (z; \, Q_0^2) \; , 
\label{e:unfcollpi+}
\\
d_1^{\bar{u}\smarrow K^+} = d_1^{d\smarrow K^+} = d_1^{s\smarrow K^+} = d_1^{\bar{d}\smarrow K^+} \equiv d_{1\, {\rm unf}}^{K^+} (z; \, Q_0^2)\; . 
\label{e:unfcollK+}
\end{gather}
The remaining favored channel $\bar{d}\to \pi^+$ is then given by 
\bea
d_1^{\bar{d}\smarrow \pi^+} &= N \, d_1^{u\smarrow \pi^+} + (N-1)\, d_1^{\bar{u}\smarrow \pi^+} \nn \\
&= N \, d_1^{u\smarrow \pi^+} + (N-1)\, d_{1\, {\rm unf}}^{\pi^+} (z; \, Q_0^2) \; .
\label{e:favcolldbarpi+}
\eea 
The $h = \pi^-$ and $h = K^-$ channels can be deduced from the above ones using the charge-conjugation symmetry of Eq.~(\ref{e:C-inv}). 

By inserting the Gaussian ansatz with the above assumptions in the expressions of Sec.~\ref{s:fav-unfav}, we get
\bea
D_{\rm fav}^{\pi^+ \pi^-} (z_1, z_2, b_{\sT}; Q_0^2) &= \[ \frac{N^2+4}{9} d_1^{u\smarrow \pi^+} (z_1; Q_0^2) \, d_1^{u\smarrow \pi^+} (z_2; Q_0^2) \right. \nn \\
&\left. \hspace{-0.5cm} + \frac{N (N-1)}{9} \, d_1^{u\smarrow \pi^+} (z_1; Q_0^2) \, d_{1\, {\rm unf}}^{\pi^+} (z_2; \, Q_0^2) + 
(1 \leftrightarrow 2) \right. \nn \\
&\left. \hspace{-1cm} + \frac{(N-1)^2}{9} d_{1\, {\rm unf}}^{\pi^+} (z_1; \, Q_0^2) \, d_{1\, {\rm unf}}^{\pi^+} (z_2; \, Q_0^2)  \] \, G_{\rm fav} (z_1, b_{\sT}^2) \, G_{\rm fav} (z_2, b_{\sT}^2)  \; , 
\label{e:favpi+pi-Gauss} 
\\[0.1cm]
D_{\rm unf}^{\pi^+ \pi^-} (z_1, z_2, b_{\sT}; Q_0^2) &= \frac{7}{9} \, d_{1\, {\rm unf}}^{\pi^+} (z_1; \, Q_0^2) \, d_{1\, {\rm unf}}^{\pi^+} (z_2; \, Q_0^2) \, G_{\rm unf} (z_1, b_{\sT}^2) \, G_{\rm unf} (z_2, b_{\sT}^2) \; ,
\label{e:unfpi+pi-Gauss}
\\[0.3cm]
D_1^{u\smarrow \pi^+} (z, b_{\sT}; Q_0^2) &= d_1^{u\smarrow \pi^+} (z; Q_0^2) \, G_{\rm fav} (z, b_{\sT}^2) \; , 
\label{e:favpi+Gauss}
\\[0.1cm]
D_1^{\bar{d}\smarrow \pi^+} (z, b_{\sT}; Q_0^2) &= \[ N\, d_1^{u\smarrow \pi^+} (z; Q_0^2) + (N-1)\, d_{1\, {\rm unf}}^{\pi^+} (z; \, Q_0^2) \] \, G_{\rm fav} (z, b_{\sT}^2) \; , 
\label{e:favpi+Gauss2}
\\[0.1cm]
D_{\rm unf}^{\pi^+} (z, b_{\sT}; Q_0^2) &= d_{1\, {\rm unf}}^{\pi^+} (z; Q_0^2) \, G_{\rm unf} (z, b_{\sT}^2) \; ,
\label{e:unfpi+Gauss}
\\[0.3cm]
D_{\rm fav}^{K^+ K^-} (z_1, z_2, b_{\sT}; Q_0^2) &= \frac{4}{9} d_1^{u\smarrow K^+} (z_1; Q_0^2) \, d_1^{u\smarrow K^+} (z_2; Q_0^2) \, G_{uK} (z_1, b_{\sT}^2) \, G_{uK} (z_2, b_{\sT}^2) \nn \\
&\hspace{-1cm}+ \frac{1}{9} d_1^{\bar{s}\smarrow K^+} (z_1; Q_0^2)\, d_1^{\bar{s}\smarrow K^+} (z_2; Q_0^2) \, G_{sK} (z_1, b_{\sT}^2) \, G_{sK} (z_2, b_{\sT}^2) \; , 
\label{e:favK+K-Gauss}
\\[0.1cm]
D_{\rm unf}^{K^+ K^-} (z_1, z_2, b_{\sT}; Q_0^2) &= \frac{7}{9} \, d_{1\, {\rm unf}}^{K^+} (z_1; \, Q_0^2) \, d_{1\, {\rm unf}}^{K^+} (z_2; \, Q_0^2) \, G_{\rm unf} (z_1, b_{\sT}^2) \, G_{\rm unf} (z_2, b_{\sT}^2) \; , 
\label{e:unfK+K-Gauss}
\\[0.3cm]
D_1^{u\smarrow K^+} (z, b_{\sT}; Q_0^2) &= d_1^{u\smarrow K^+} (z; Q_0^2) \, G_{uK} (z, b_{\sT}^2) \; , 
\label{e:favK+Gauss}
\\[0.1cm]
D_1^{\bar{s}\smarrow K^+} (z, b_{\sT}; Q_0^2) &\equiv D_1^{s\smarrow K^-} (z, b_{\sT}; Q_0^2) = d_1^{\bar{s}\smarrow K^+} (z; Q_0^2) \, G_{sK} (z, b_{\sT}^2) \; , 
\label{e:favK+Gauss2}
\\[0.1cm]
D_{\rm unf}^{K^+} (z, b_{\sT}; Q_0^2) &= d_{1\, {\rm unf}}^{K^+} (z; \, Q_0^2) \, G_{\rm unf} (z, b_{\sT}^2) \; ,
\label{e:unfK+Gauss}
\\[0.1cm]
D_{\rm fav}^{\pi^+ K^-} (z_1, z_2, b_{\sT}; Q_0^2) &= \frac{4}{9} d_1^{u\smarrow \pi^+} (z_1; Q_0^2) \, d_1^{u\smarrow K^+} (z_2; Q_0^2) \, G_{\rm fav} (z_1, b_{\sT}^2) \, G_{uK} (z_2, b_{\sT}^2) \; .
\label{e:favpi+K-Gauss}
\eea

\section{Predictions for TMD multiplicities}
\label{s:results}

In this section, we present our results as normalized multiplicities 
\beq
M^{h_1 h_2} (z_1, z_2, q_{\sT}^2, y) / M^{h_1 h_2} (z_1, z_2, 0, y) 
\label{e:norm_mult}
\eeq
for the hadron pair $(h_1, h_2)$, where $M^{h_1 h_2} (z_1, z_2, q_{\sT}^2, y)$ is defined in Eq.~(\ref{e:multi}). In such way, we are able to directly compare the genuine trend in $q_{\sT}^2$ for each different case. If not explicitly specified, we choose $y = 0.2$. For selected values of $\{ z_1, \, z_2\}$, the results are displayed as a function of $\bm{P}_{1\p}^2 = z_1^2 \bm{q}_{\sT}^2$. Hence, the useful range in $\bm{P}_{1\p}^2$ depends on $z_1$ in order to fulfill the condition $q_{\sT}^2 \ll Q^2$. The range obviously depends also on the choice of the hard scale; we consider $Q^2 = 100$ GeV$^2$, as in the \belle experiment, and $Q^2 = 14.6$ GeV$^2$, as in the \bes one. For each specific case, the results are displayed as uncertainty bands: they represent the 68\% of the envelope of 200 different values for the intrinsic parameters in Eqs.~(\ref{e:width})-(\ref{e:unfwidth}) for the $D_1 (z, b_{\sT}; Q_0^2)$ at the starting scale $Q_0^2$, obtained by rejecting the largest and lowest 16\% of them. The 200 values are obtained by fitting 200 replicas of SIDIS multiplicities measured by the \hermes collaboration~\cite{Airapetian:2012ki}. If the 200 values for each parameter were distributed as a Gaussian, the 68\% band would correspond to the usual $1\sigma$ confidence interval (for more details, see Ref.~\cite{Signori:2013mda}).

The results are organized as follows. In Sec.~\ref{ss:evo_param}, we show the sensitivity of the normalized multiplicity to different values of the evolution parameters $\{\bmax, g_2\}$ described  in Sec.~\ref{ss:collins_evo} for a final hadron pair $(h_1 h_2) = (\pi^+ \pi^-)$. In Sec.~\ref{ss:evo_scheme}, we compare normalized multiplicities for the two different  evolution schemes described in Secs.~\ref{ss:collins_evo} and~\ref{ss:fixedscale_evo}. In Sec.~\ref{ss:matching}, we discuss the capability of discriminating among the various prescriptions illustrated in Sec.~\ref{ss:collins_evo} for the nonperturbative evolution effects. In Sec.~\ref{ss:z-trend}, we concentrate on the sensitivity of the normalized multiplicities upon varying the fractional energy $z$ of final hadrons. In Sec.~\ref{ss:BES-results}, we show how the results get modified when lowering $Q^2$ from the \belle scale to the \bes scale. Finally, in Sec.~\ref{ss:ratio-result} we discuss the sensitivity of ratios of normalized multiplicities for different final states to the flavor structure of the intrinsic transverse-momentum-dependent part of the input TMD FF at the starting scale of evolution.

\subsection{Sensitivity to nonperturbative evolution parameters}
\label{ss:evo_param}

As already remarked in Sec.~\ref{ss:collins_evo}, for a specific evolution scheme the nonperturbative part of the TMD evolution depends on the choice of a prescription for describing the transition from perturbative to nonperturbative regimes, which in turn depends on the two parameters $\bmax$ and $g_2$. In this section, we explore the sensitivity of our predictions to different values of the pair $\{ \bmax, \, g_2\}$. We adopt as limiting cases the choices $\{ \bmax = 1.5, \, g_2 = 0.18\}$ and $\{ \bmax = 0.5, \, g_2 = 0.68\}$, that were deduced in Refs.~\cite{Konychev:2005iy} and~\cite{Landry:2002ix}, respectively, by fitting the transverse-momentum distribution of lepton pairs produced in Drell-Yan processes. If not explicitly specified, the first choice is described by uncertainty bands with dot-dashed borders while the second choice is linked to bands with solid borders. As explained in Sec.~\ref{ss:collins_evo}, the two parameters are anticorrelated. In the following, we show results also for the interpolating choice $\{ \bmax = 1, \, g_2 = 0.43\}$. The corresponding results are displayed as uncertainty bands with dashed borders. 

\bef
 \includegraphics[width=0.6\textwidth]{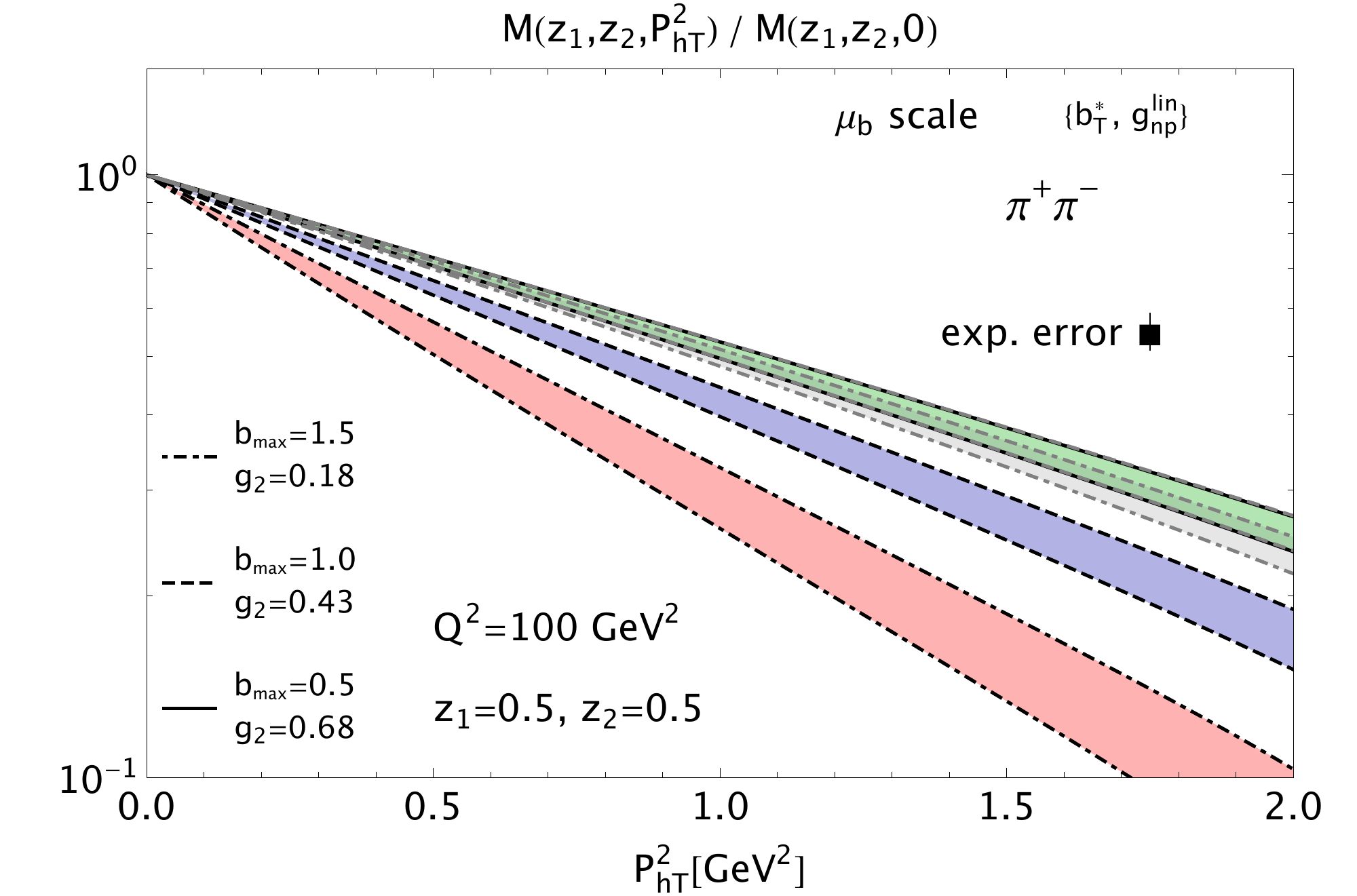}
  \caption{The normalized multiplicity at $z_1=z_2=0.5$ as a function of $\bm{P}_{1\p}^2 = z_1^2 \bm{q}_{\sT}^2 \equiv (0.5)^2 \bm{q}_{\sT}^2$ at the \belle scale $Q^2=100$ GeV$^2$ for the "$\mu_b$ scale" evolution scheme and with the $\{b_T^*, \, g_{\text{np}}^{\text{lin}}\}$ prescription for the transition to the nonperturbative regime (see text). The uncertainty bands correspond to various choices of the nonperturbative parameters of evolution: $\{ \bmax = 1.5, \, g_2 = 0.18\}$ for the band with dot-dashed borders, $\{ \bmax = 1, \, g_2 = 0.43\}$ for the one with dashed borders, $\{ \bmax = 0.5, \, g_2 = 0.68\}$ for the one with solid borders. The latter is accompanied by a light-gray band with dot-dashed borders, that represents the result with the same parameters but with the choice $\mu_b / 2$ for the arbitrary matching scale, and by an overlapping light-gray band with dashed borders for the choice $2 \mu_b$. An experimental error of 7\% is also indicated.}
  \label{f:bmaxg2}
\eef

In Fig.~\ref{f:bmaxg2}, the normalized multiplicity 
\beq
M^{\pi^+ \pi^-} (z_1=0.5, z_2=0.5, q_{\sT}^2, y=0.2) / M^{\pi^+ \pi^-} (z_1=0.5, z_2=0.5, 0, y=0.2)
\label{e:norm_mult_fig2}
\eeq
is shown as a function of $\bm{P}_{1\p}^2 = z_1^2 \bm{q}_{\sT}^2 \equiv (0.5)^2 \bm{q}_{\sT}^2$ at the \belle scale $Q^2=100$ GeV$^2$ for the "$\mu_b$ scale" evolution scheme and with the $\{b_T^*, \, g_{\text{np}}^{\text{lin}}\}$ prescription for the transition to the nonperturbative regime, as explained in Sec.~\ref{ss:collins_evo}. The explored range in $\bm{P}_{1\p}^2$ is such that for $z_1=0.5$ the maximum $\bm{q}_{\sT}^2$ satisfies the condition $\bm{q}_{\sT}^2 \ll Q^2$. The three uncertainty bands, corresponding to the three different choices $\{ \bmax = 1.5, \, g_2 = 0.18\}$ (dot-dashed borders), $\{ \bmax = 1, \, g_2 = 0.43\}$ (dashed borders), and $\{ \bmax = 0.5, \, g_2 = 0.68\}$ (solid borders), are well separated. The squared box with error bar indicates a hypothetical experimental error of 7\%. We fix it by propagating to the normalized multiplicity the typical experimental error of 3\% for  single-hadron production data in $e^+e^-$ annihilations at $Q^2 = 100$ GeV$^2$ and $z = 0.5$, from which the collinear $d_1^q (z; \, Q^2)$ are extracted~\cite{deFlorian:2007aj}. This experimental error of 7\% seems small enough to discriminate among predictions produced with different choices of $\{ \bmax, \, g_2\}$. 

Two additional light-gray bands are shown, which are partially overlapped (dot-dashed borders) or completely overlapped (dashed borders) to the band with solid borders corresponding to the choice $\{ \bmax = 0.5, \, g_2 = 0.68\}$. These bands reproduce the outcome of calculations performed in the same conditions but for different (arbitrary) choices of the scale $\mb$. If the band with solid borders corresponds to calculations with the choice of Eq.~(\ref{e:def_mub}) for $\mb$, then the light-gray band with dot-dashed borders corresponds to the choice $\mb / 2$, and the one with dashed borders to $2 \mb$. The almost complete overlap of these results shows that for the selected observable, the normalized multiplicity, the theoretical uncertainty in determining the matching scale $\mb$ (that describes the transition from perturbative to nonperturbative regimes) is negligible with respect to the sensitivity to the parameters describing the nonperturbative effects in the evolution. 

\subsection{Sensitivity to evolution schemes}
\label{ss:evo_scheme}

In this section, we explore the sensitivity of our normalized multiplicity to the choice of the evolution scheme. In Sec.~\ref{s:evolution}, we described two different schemes, the "$\mb$ scale" and the "fixed scale". They differ mainly in the fact that in the latter the whole distribution in impact parameter space $b_{\sT}$ of the TMD FF $D_1^q$ at beginning of evolution is computed at a fixed scale $Q_0$, namely there is no impact parameter that describes the transition from low (perturbative) $b_{\sT}$ to high (nonperturbative) $b_{\sT}$. Actually, one would expect that for small values of $g_2$ and corresponding not too large values of $\bmax$ ({\it i.e.}, where the perturbative description of the evolution of the $b_{\sT}$ distribution is still applicable and gives the predominant contribution) the predictions from the different schemes should tend to a common result, determined mainly by a fully perturbative calculation. However, the complexity of the evolution kernels, described in Secs.~\ref{ss:collins_evo} and~\ref{ss:fixedscale_evo}, indicates that this is too a na\"ive expectation. 

\bef
 \includegraphics[width=0.6\textwidth]{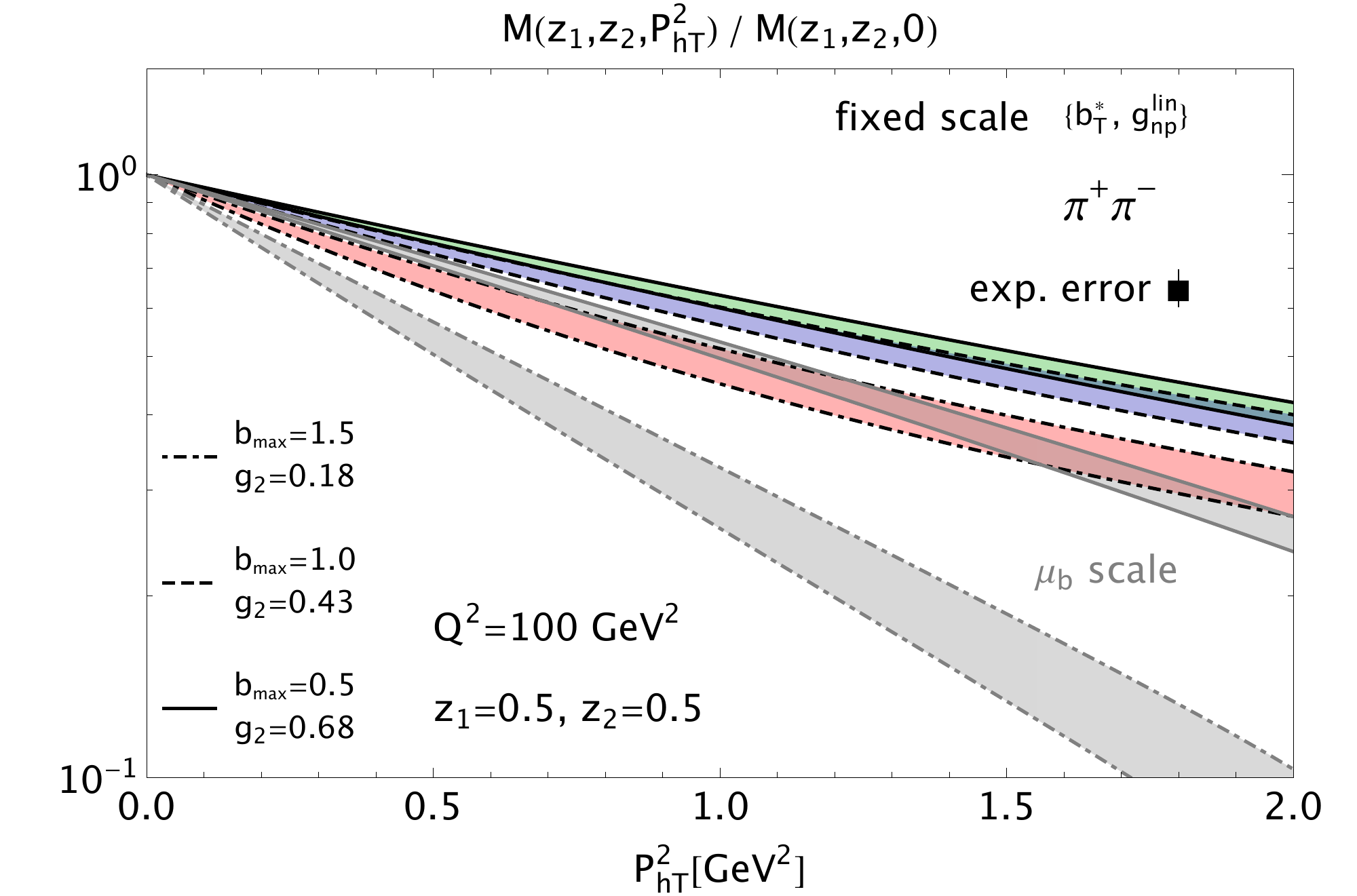}
  \caption{The normalized multiplicity at $z_1=z_2=0.5$ as a function of $\bm{P}_{1\p}^2 = z_1^2 \bm{q}_{\sT}^2 \equiv (0.5)^2 \bm{q}_{\sT}^2$ in the same conditions and with the same notation as in Fig.~\ref{f:bmaxg2}, but for the "fixed scale" evolution scheme. The additional light-gray bands with dot-dashed and solid borders are the result related to the "$\mu_b$ scale" evolution scheme for $\{ \bmax = 1.5, \, g_2 = 0.18\}$ and $\{ \bmax = 0.5, \, g_2 = 0.68\}$, respectively.}
  \label{f:mub-fixed}
\eef

In fact, in Fig.~\ref{f:mub-fixed} the normalized multiplicity of Eq.~(\ref{e:norm_mult_fig2}) is shown as a function of $\bm{P}_{1\p}^2 = z_1^2 \bm{q}_{\sT}^2 \equiv (0.5)^2 \bm{q}_{\sT}^2$ at the \belle scale $Q^2=100$ GeV$^2$ with the $\{b_T^*, \, g_{\text{np}}^{\text{lin}}\}$ prescription. There are two groups of uncertainty bands. The former one displays the results for the "fixed scale" evolution scheme in the standard notation, {\it i.e.} for $\{ \bmax = 1.5, \, g_2 = 0.18\}$ (dot-dashed borders), $\{ \bmax = 1, \, g_2 = 0.43\}$ (dashed borders), and $\{ \bmax = 0.5, \, g_2 = 0.68\}$ (solid borders). Then, two additional light-gray bands are shown that correspond to the results with the "$\mb$ scale" evolution scheme for $\{ \bmax = 1.5, \, g_2 = 0.18\}$ (dot-dashed borders) and $\{ \bmax = 0.5, \, g_2 = 0.68\}$ (solid borders). 

It is evident that for the maximum (minimum) $\bmax$ ($g_2$) the band with dot-dashed borders in the "fixed scale" scheme is not similar to the light-gray band with dot-dashed borders in the "$\mb$ scale" scheme. Actually, all the results in the "fixed scale" scheme show a much larger distribution in $\bm{P}_{1\p}^2$, somewhat pointing to stronger evolution effects of perturbative origin that seem to be absent in the "$\mb$ scale" scheme (where the scale choice minimizes the effect of large logarithms in the perturbative coefficients). It is important to notice that there is a significant overlap between the band with dot-dashed borders in the "fixed scale" scheme and the light-gray band with solid borders in the "$\mb$ scale" scheme. Apparently, the normalized multiplicity seems not to be enough sensitive to discriminate among different evolution schemes, since two different choices of them can produce similar results with different evolution parameters $\{ \bmax, \, g_2\}$. However, this result is observed at a specific value of fractional energies of the final hadrons, namely $z_1 =  z_2 = 0.5$. 

\bef
 \includegraphics[width=0.6\textwidth]{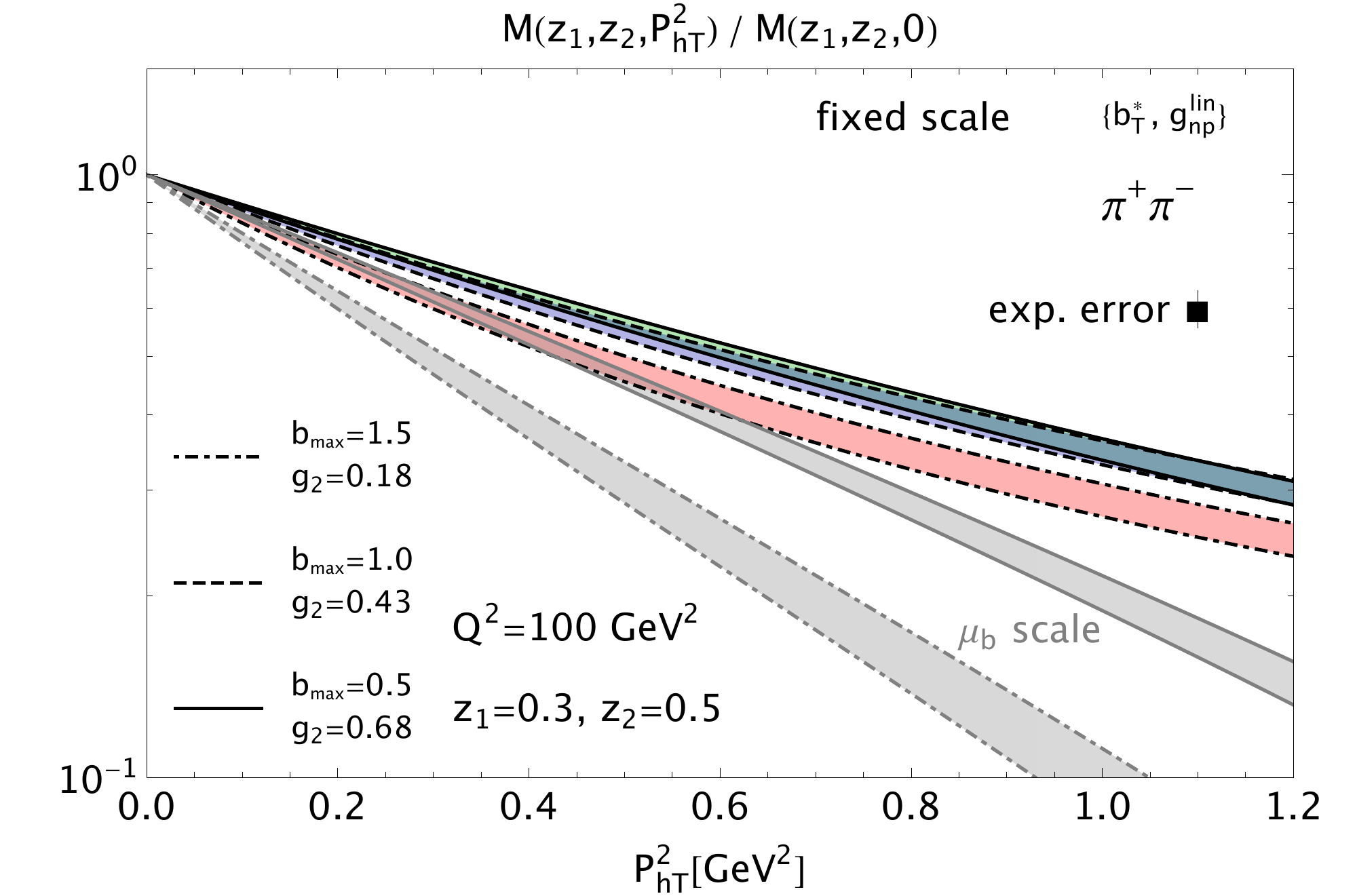}
  \caption{The same as in the previous figure, but at $z_1 = 0.3$.}
  \label{f:mub-fixed-lowz}
\eef

In Fig.~\ref{f:mub-fixed-lowz}, we show the $\bm{P}_{1\p}^2$ distribution of normalized multiplicities calculated in the same conditions, notation and conventions as in the previous figure, but at $z_1 = 0.3$ and $z_2 = 0.5$. The band with dot-dashed borders in the "fixed scale" scheme can now be easily separated from the light-gray band with solid borders in the "$\mb$ scale" scheme if the indicated hypothetical experimental error is around 7\%. Therefore, only when combining the study of both the $z$ and $\bm{P}_{1\p}^2$ dependencies in the normalized multiplicity we may be able to discriminate among different TMD evolution schemes. 

\subsection{Sensitivity to prescriptions for the transition to nonperturbative transverse momenta}
\label{ss:matching}

We now focus on exploring the possibility of discriminating among different prescriptions that describe the functional dependence in $b_{\sT}$ of the nonperturbative Sudakov evolution factor (see Eqs.~(\ref{e:lin_model}) and~(\ref{e:log_model})) or the transition from the perturbative low$-b_{\sT}$ domain to the nonperturbative high$-b_{\sT}$ one (see Eqs.~(\ref{e:bstar}) and (\ref{e:bsun})). 

\bef
 \includegraphics[width=0.6\textwidth]{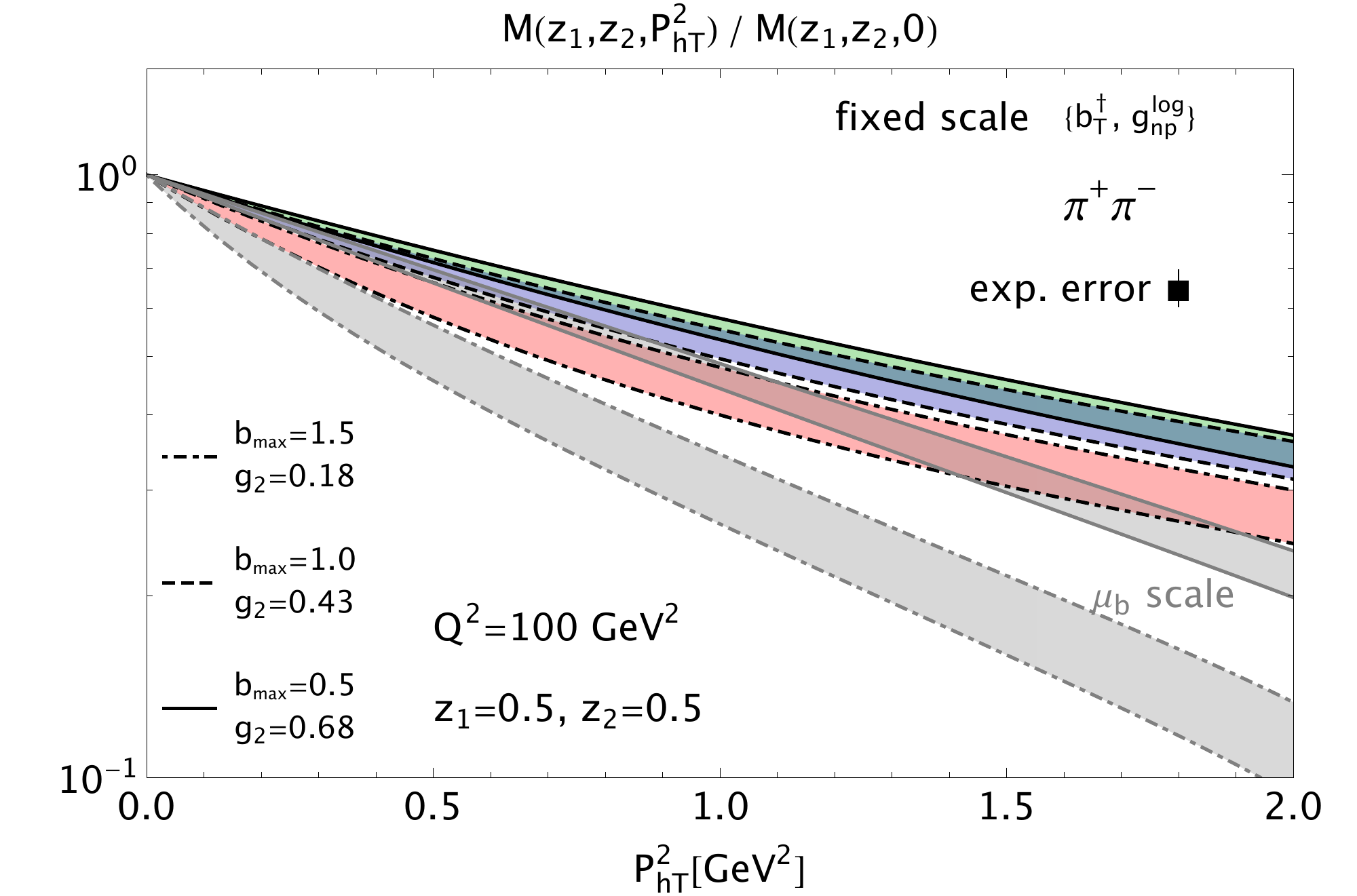}
  \caption{The normalized multiplicity at $z_1=z_2=0.5$ as a function of $\bm{P}_{1\p}^2 = z_1^2 \bm{q}_{\sT}^2 \equiv (0.5)^2 \bm{q}_{\sT}^2$ at the \belle scale $Q^2=100$ GeV$^2$ for the "fixed  scale" evolution scheme and with the $\{b_T^{\dagger}, \, g_{\text{np}}^{\text{log}}\}$ prescription for the transition to the nonperturbative regime (see text). Notation and conventions for the uncertainty bands as in Fig.~\ref{f:mub-fixed}.}
  \label{f:expLog}
\eef

In Fig.~\ref{f:expLog}, the normalized multiplicity of Eq.~(\ref{e:norm_mult_fig2}) is shown as a function of $\bm{P}_{1\p}^2 = z_1^2 \bm{q}_{\sT}^2 \equiv (0.5)^2 \bm{q}_{\sT}^2$ at the \belle scale $Q^2=100$ GeV$^2$ with the $\{b_T^{\dagger}, \, g_{\text{np}}^{\text{log}}\}$ prescription. Again, as in Fig.~\ref{f:mub-fixed} there are two groups of uncertainty bands. The former one displays the results for the "fixed scale" evolution scheme in the standard notation, {\it i.e.} for $\{ \bmax = 1.5, \, g_2 = 0.18\}$ (dot-dashed borders), $\{ \bmax = 1, \, g_2 = 0.43\}$ (dashed borders), and $\{ \bmax = 0.5, \, g_2 = 0.68\}$ (solid borders). The two additional light-gray bands correspond to the results with the "$\mb$ scale" evolution scheme for $\{ \bmax = 1.5, \, g_2 = 0.18\}$ (dot-dashed borders) and $\{ \bmax = 0.5, \, g_2 = 0.68\}$ (solid borders). So, also for the $\{b_T^{\dagger}, \, g_{\text{np}}^{\text{log}}\}$ prescription we find the same ambiguity as for the $\{b_T^*, \, g_{\text{np}}^{\text{lin}}\}$ one in Fig.~\ref{f:mub-fixed}: the overlap of the light-gray band with solid borders and of the band with dot-dashed borders indicates that two different evolution schemes give similar results with different evolution parameters $\{ \bmax, \, g_2\}$. Hence, we wonder if this similar trend suggests that it might not be possible to distinguish between the two schemes. Again, the possible way out is to look at the dependence of the results upon the fractional energy of the final hadrons. 

\bef
 \includegraphics[width=0.4\textwidth]{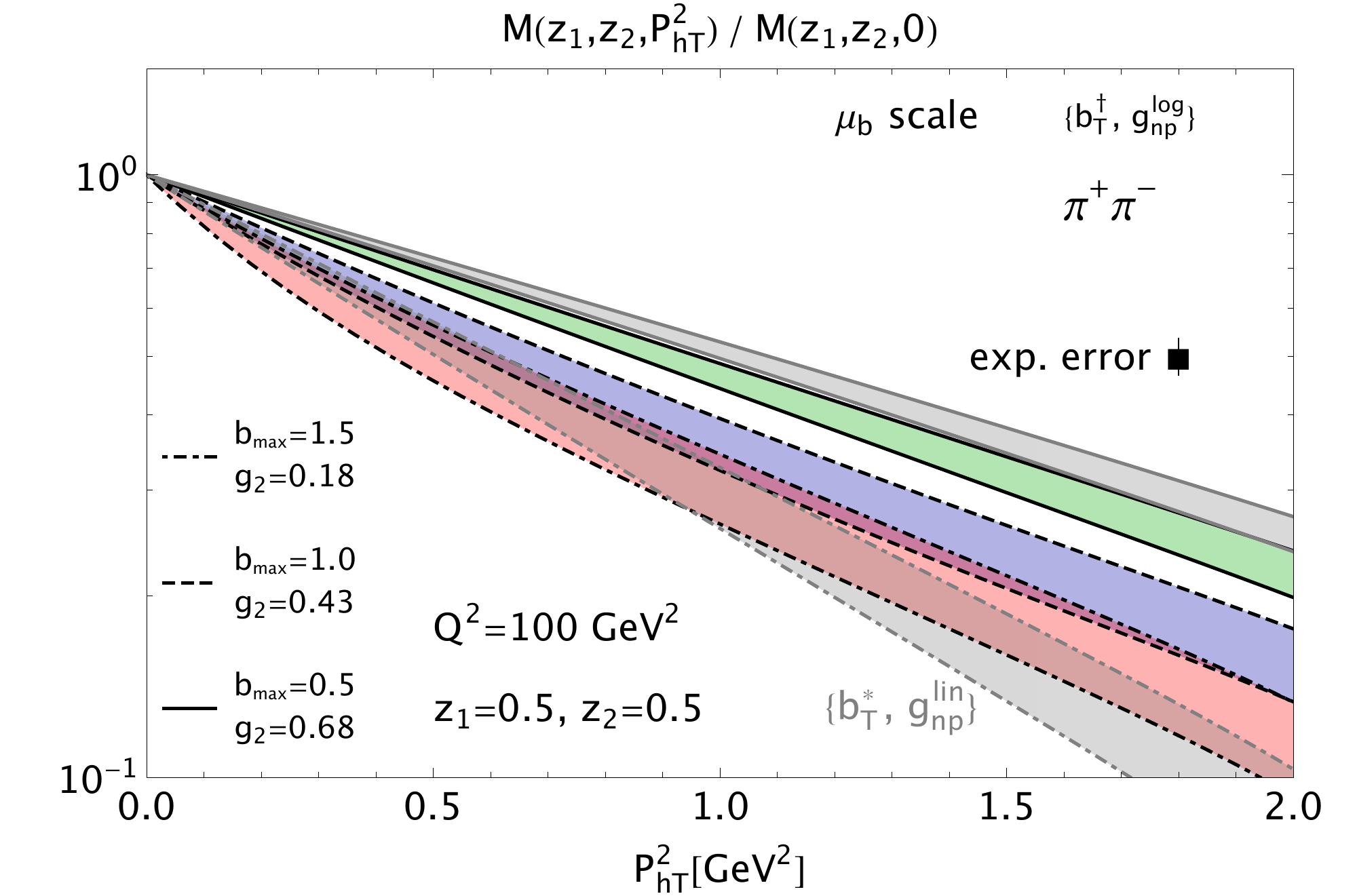}\hspace{.5cm} \includegraphics[width=0.4\textwidth]{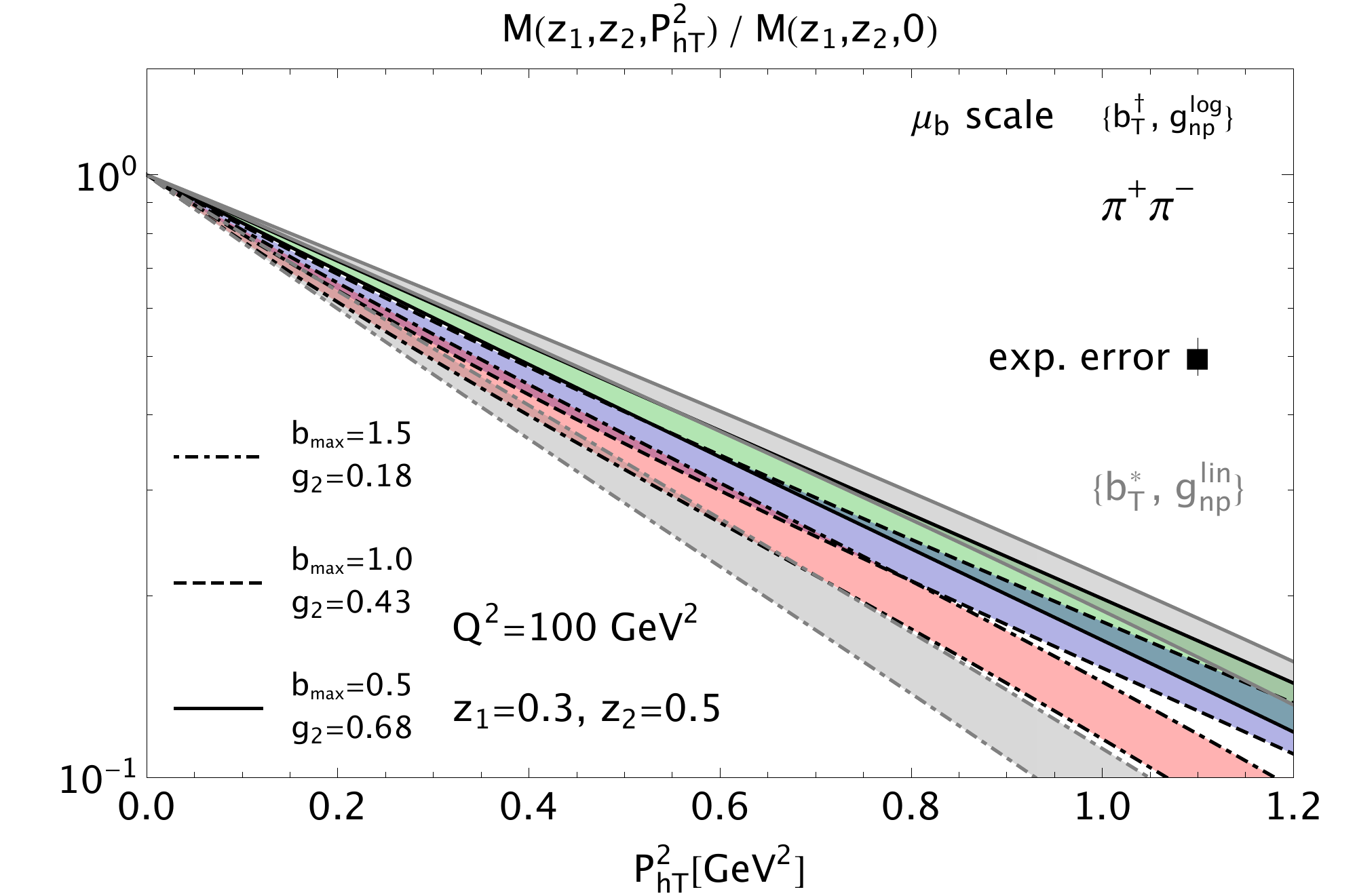}
  \caption{The normalized multiplicity at $z_2=0.5$ as a function of $\bm{P}_{1\p}^2 = z_1^2 \bm{q}_{\sT}^2$ at the \belle scale $Q^2=100$ GeV$^2$ for the "$\mu_b$ scale" evolution scheme and with the $\{b_T^{\dagger}, \, g_{\text{np}}^{\text{log}}\}$ prescription for the transition to the nonperturbative regime (see text). Notation for the uncertainty bands as in previous figure. The additional light-gray bands with dot-dashed and solid borders are the result with the $\{b_T^*, \, g_{\text{np}}^{\text{lin}}\}$ matching prescription for $\{ \bmax = 1.5, \, g_2 = 0.18\}$ and $\{ \bmax = 0.5, \, g_2 = 0.68\}$, respectively. Left panel for $z_1=0.5$, right panel for $z_1=0.3$.}
  \label{f:b*-blog}
\eef

In Fig.~\ref{f:b*-blog}, the normalized multiplicity of Eq.~(\ref{e:norm_mult}) is shown as a function of $\bm{P}_{1\p}^2 = z_1^2 \bm{q}_{\sT}^2$ at the \belle scale $Q^2=100$ GeV$^2$ for the "$\mb$ scale" evolution scheme. Also in this plot, there are two groups of uncertainty bands. A group displays the results for the  $\{b_T^{\dagger}, \, g_{\text{np}}^{\text{log}}\}$ prescription in the standard notation, {\it i.e.} for $\{ \bmax = 1.5, \, g_2 = 0.18\}$ (dot-dashed borders), $\{ \bmax = 1, \, g_2 = 0.43\}$ (dashed borders), and $\{ \bmax = 0.5, \, g_2 = 0.68\}$ (solid borders). The group of two light-gray bands correspond to the results with the $\{b_T^*, \, g_{\text{np}}^{\text{lin}}\}$ prescription for $\{ \bmax = 1.5, \, g_2 = 0.18\}$ (dot-dashed borders) and $\{ \bmax = 0.5, \, g_2 = 0.68\}$ (solid borders). If we focus on the left panel where calculations are performed at $z_1=z_2=0.5$, the two bands with dot-dashed borders are substantially overlapped, thus reinforcing the suspect that it might not be possible to discriminate between the $\{b_T^*, \, g_{\text{np}}^{\text{lin}}\}$ and $\{b_T^{\dagger}, \, g_{\text{np}}^{\text{log}}\}$ prescriptions. But if we now turn to the right panel, where the same calculation is performed at $z_1=0.3, \, z_2=0.5$, we may hope to have a sufficiently small experimental error that discriminates between the two bands with dot-dashed borders. Unfortunately, the plot suggests also that this option seems possible only for the $\{ \bmax = 1.5, \, g_2 = 0.18\}$ case. And further explorations show that the same calculation, when performed in the "fixed scale" evolution scheme, produces more confused results. In summary, a combined study of the $z$ and $\bm{P}_{1\p}^2$ dependencies in the normalized multiplicity might be able to discriminate among different prescriptions for the nonperturbative effects in the evolution only for a selected set of evolution parameters and schemes.

\subsection{Sensitivity to hadron fractional-energy dependence}
\label{ss:z-trend}

In the previous sections, we found that in several occasions only the combined study of the $z$ and $\bm{P}_{1\p}^2$ dependencies of the normalized multiplicity allows for discerning results obtained from different parametrizations and prescriptions in the description of nonperturbative effects in the TMD evolution. This is not accidental. With the approximations adopted in this work, the main difference between the two considered evolution schemes lies in fact in the $z$ dependence of the collinear fragmentation function $d_1$, as it can be deduced by comparing Eqs.~(\ref{e:nCSS}) and~(\ref{e:hybrid}). 

\bef
 \includegraphics[width=0.4\textwidth]{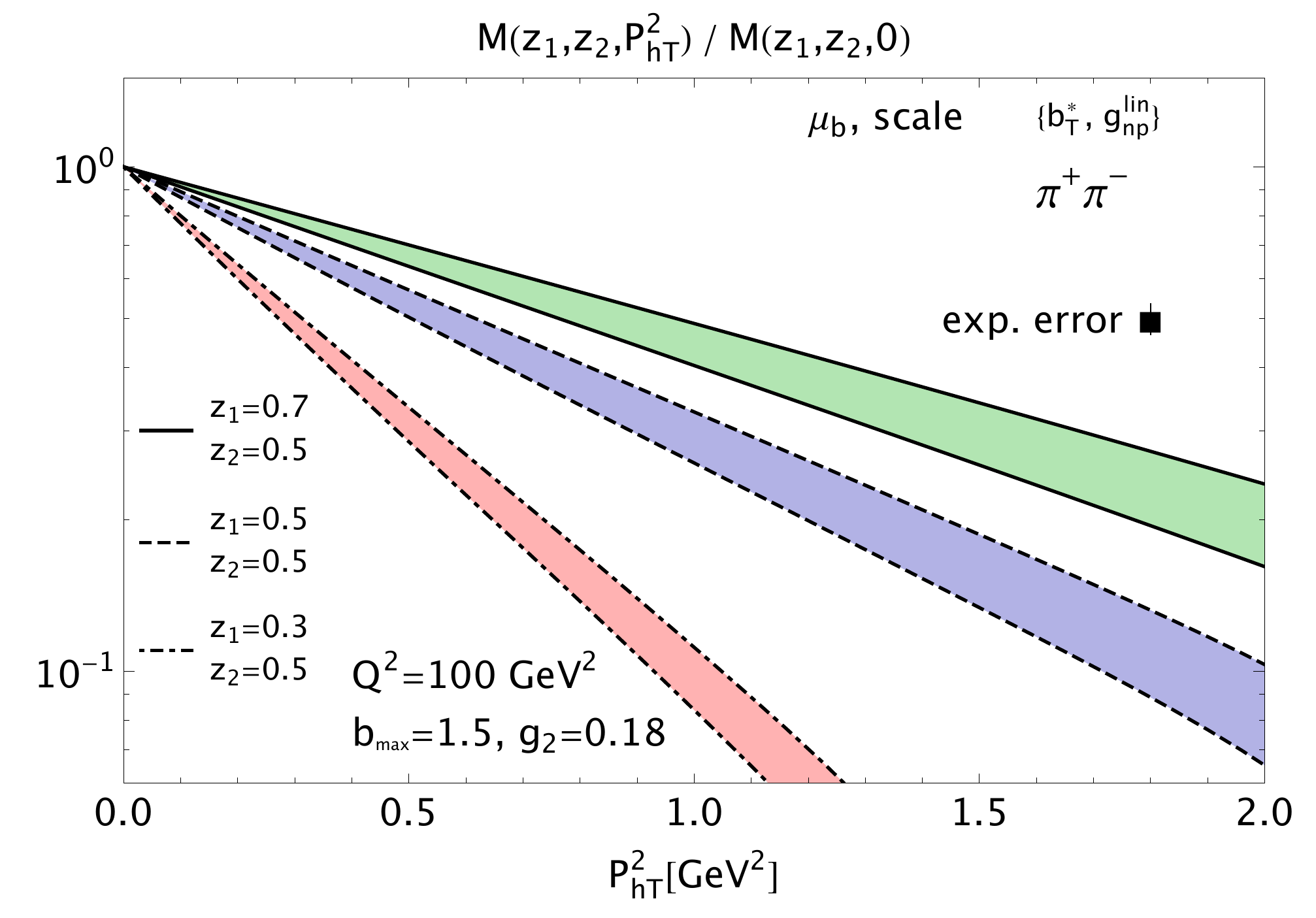}\hspace{.5cm} \includegraphics[width=0.4\textwidth]{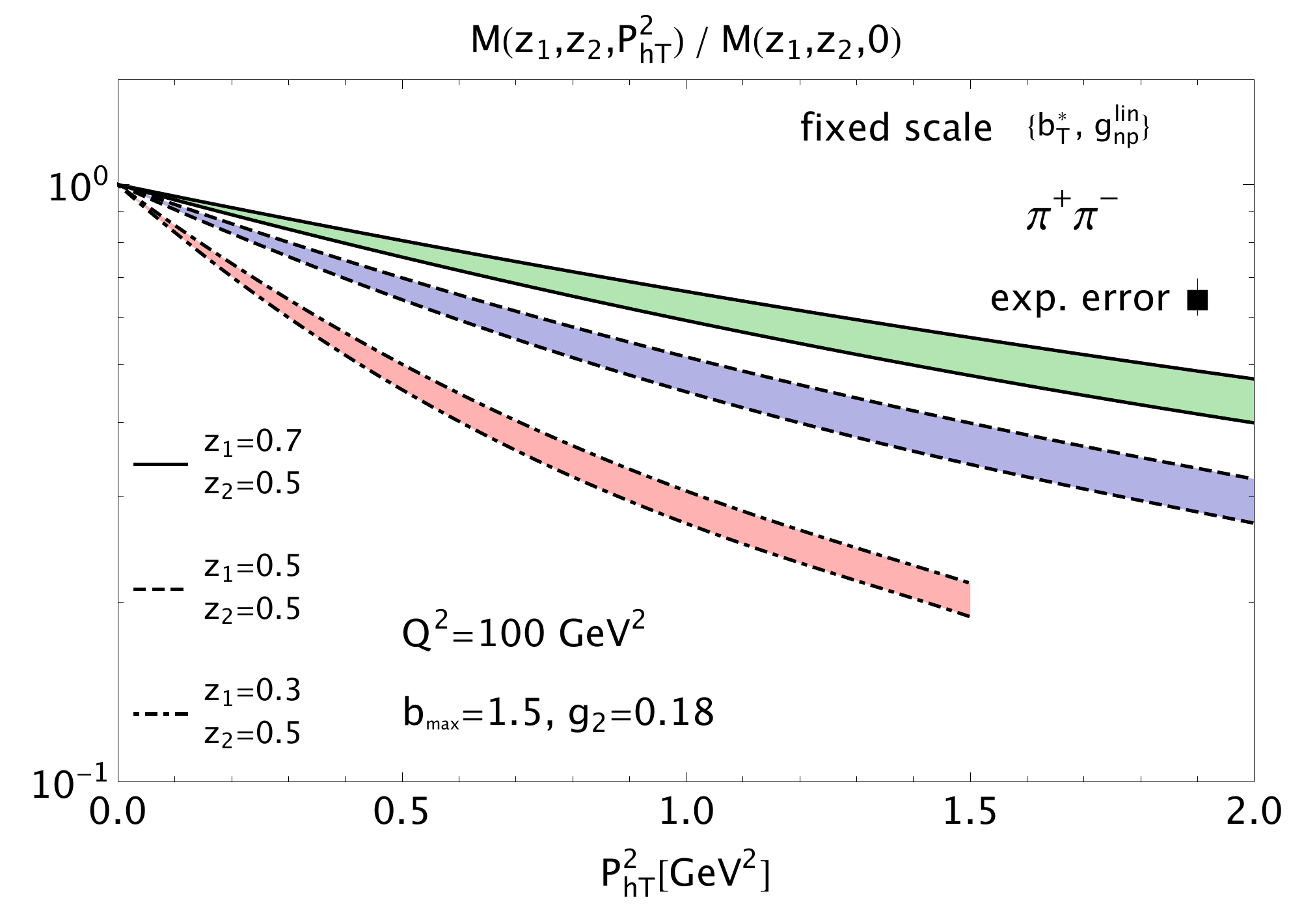}
  \caption{The normalized multiplicity at $z_2=0.5$ as a function of $\bm{P}_{1\p}^2 = z_1^2 \bm{q}_{\sT}^2$ at the \belle scale $Q^2=100$ GeV$^2$ for the evolution parameters $\{ \bmax = 1.5, \, g_2 = 0.18\}$ and with the $\{b_T^*, \, g_{\text{np}}^{\text{lin}}\}$ prescription for the transition to the nonperturbative regime (see text). Uncertainty band with dot-dashed borders for $z_1=0.3$, with dashed borders for $z_1=0.5$, with solid borders for $z_1=0.7$. The squared box with error bar corresponds to an experimental error of 7\%. Left panel for the "$\mu_b$ scale" evolution scheme, right panel for the "fixed scale" one.}
  \label{f:z-mub-fixed}
\eef

The plots in Fig.~\ref{f:z-mub-fixed} seem to confirm this finding. In the left panel, the normalized multiplicity of Eq.~(\ref{e:norm_mult}) is shown as a function of $\bm{P}_{1\p}^2 = z_1^2 \bm{q}_{\sT}^2$ at the \belle scale $Q^2=100$ GeV$^2$ for the "$\mb$ scale" evolution scheme, the $\{ b_{\sT}^*, \, g_{\text{np}}^{\text{lin}}\}$ prescription, and the choice $\{ \bmax = 1.5, \, g_2 = 0.18\}$. The bands display results for the values $z_1=0.3, \, z_2=0.5$ (band with dot-dashed borders), $z_1=z_2=0.5$ (dashed borders), and $z_1=0.7, \, z_2=0.5$ (solid borders). In the right panel, we show the results of the calculations performed in the same conditions but for the "fixed scale" evolution scheme. It is quite evident that the latter scheme produces $\bm{P}_{1\p}^2$ distributions that are systematically larger for any combination of $\{ z_1, z_2\}$. This finding holds true also for other choices of the evolution parameters $\{ \bmax, \, g_2\}$ and for the $\{b_T^{\dagger}, \, g_{\text{np}}^{\text{log}}\}$ prescription.

\subsection{Sensitivity to the hard scale: from \belle to \bes}
\label{ss:BES-results}

All previous results have been obtained at the \belle scale of $Q^2 = 100$ GeV$^2$. We may wonder what happens when reducing the "evolution path" to lower scales, like, e.g., the \bes scale $Q^2 = 14.6$ GeV$^2$. 

\bef
 \includegraphics[width=0.6\textwidth]{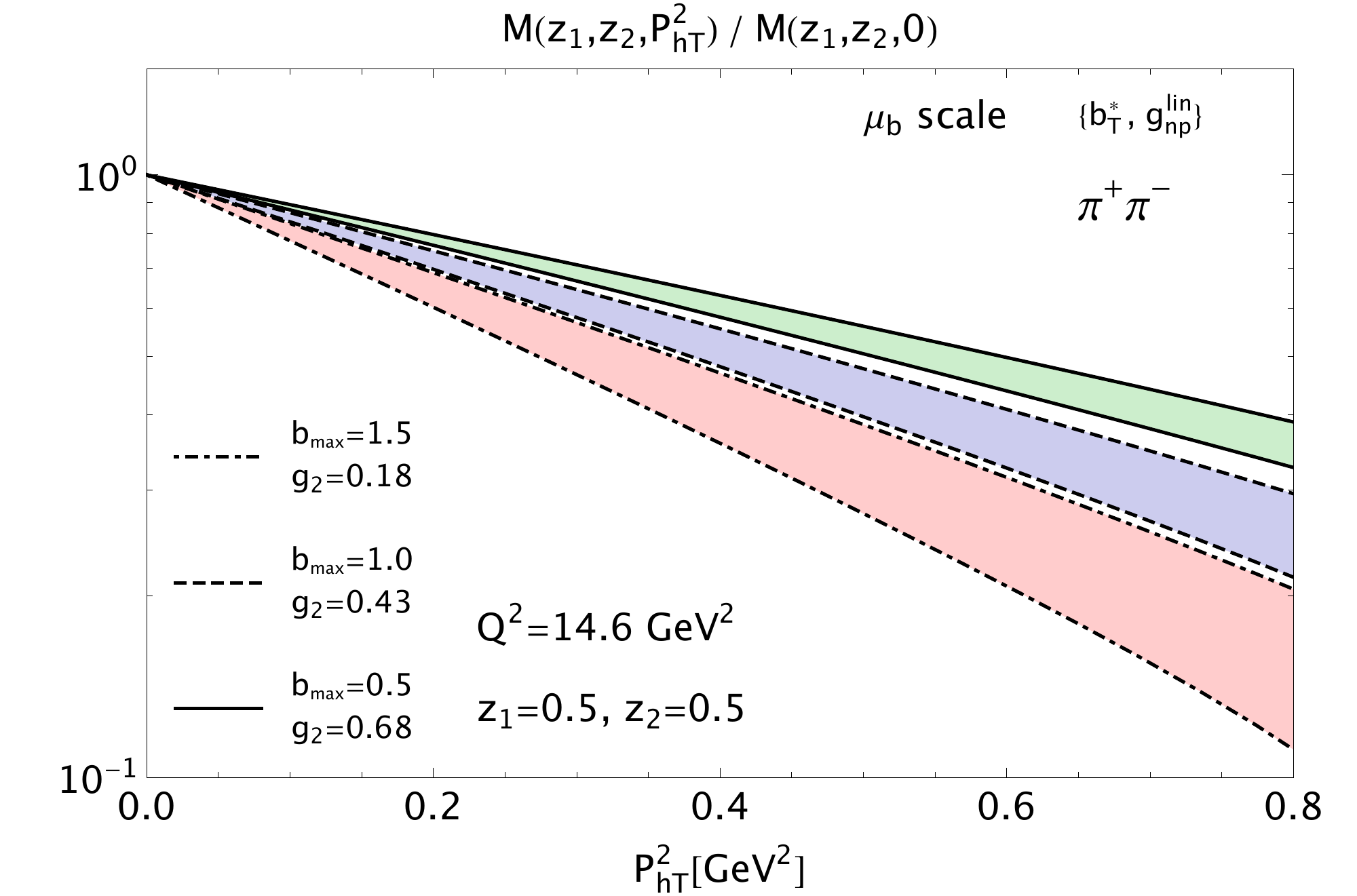}
  \caption{The normalized multiplicity of Eq.~(\ref{e:norm_mult_fig2}) as a function of $\bm{P}_{1\p}^2 = z_1^2 \bm{q}_{\sT}^2 \equiv (0.5)^2 \bm{q}_{\sT}^2$ at the \bes scale $Q^2=14.6$ GeV$^2$ for the "$\mu_b$ scale" evolution scheme and with the $\{b_T^*, \, g_{\text{np}}^{\text{lin}}\}$ prescription for the transition to the nonperturbative regime (see text). Notation and conventions for the uncertainty bands as in Fig.~\ref{f:bmaxg2}.}
  \label{f:BES}
\eef

In Fig.~\ref{f:BES}, the normalized multiplicity of Eq.~(\ref{e:norm_mult_fig2}) is shown as a function of $\bm{P}_{1\p}^2 = z_1^2 \bm{q}_{\sT}^2 \equiv (0.5)^2 \bm{q}_{\sT}^2$ in the same conditions and notation as in Fig.~\ref{f:bmaxg2} but at the \bes scale $Q^2 = 14.6$ GeV$^2$. By comparing these results with the ones in Fig.~\ref{f:bmaxg2}, we deduce that the net effect is a systematic enlargement of the uncertainty bands. This finding occurs also for other combinations of evolutions schemes and nonperturbative prescriptions. Hence, we deduce that working at the \bes scale is not useful if we want to discriminate among different evolution parameters $\{ \bmax, \, g_2\}$, or between the $\{ b_{\sT}^*, \, g_{\text{np}}^{\text{lin}}\}$ and $\{b_T^{\dagger}, \, g_{\text{np}}^{\text{log}}\}$ prescriptions, or between the "fixed scale" and "$\mb$ scale" evolution schemes. 

However, we recall that each uncertainty band is the envelope of the 68\% of 200 different curves, each one corresponding to a specific replica of the intrinsic parameters entering the Gaussian widths $\langle \bm{P}^2_{\p}\rangle^{a\smarrow h} (z)$ of Eq.~(\ref{e:TMDFFQ0}) for the $b_{\sT}$ distribution of the $D_1^a$ at the starting scale in the evolution. Then, we might envisage that the experimental error is sufficiently smaller than the band width such that it is able to discriminate some of the replicas, in order to narrow the uncertainty on the intrinsic parameters. In any case, this goal will be achieved only by performing additional more precise measurements of SIDIS multiplicities for different final hadron species and on different targets. 

\subsection{Sensitivity to partonic flavor}
\label{ss:ratio-result}

The sensitivity to the nonperturbative intrinsic parameters, that describe the $b_{\sT}$ distribution of the TMD FF at the initial scale of evolution, is an important issue. The analysis of SIDIS multiplicities at low $Q^2$ suggests that some of these parameters are different for different flavors~\cite{Signori:2013mda}. Hence, we expect that also the distribution in transverse momentum space of the evolved TMD FF will  depend on the flavor of the fragmenting partons. However, the cross section in Eq.~(\ref{e:xsect}) mixes all flavors in the sum. Therefore, it is useful to define an observable that is well suited to explore the effect of flavor in the TMD evolution. 

In the following, we will show results for the $\bm{P}_{1\p}^2$ distribution of ratios of normalized multiplicities corresponding to different final states:
\beq
M^{h_1 h_2} (z_1, z_2, q_{\sT}^2, y) / M^{h_1 h_2} (z_1, z_2, 0, y) \, \times \, \left[ 
M^{h_1' h_2'} (z_1, z_2, q_{\sT}^2, y) / M^{h_1' h_2'} (z_1, z_2, 0, y) \right]^{-1} \; .
\label{e:ratio_norm_mult}
\eeq

\bef
 \includegraphics[width=0.4\textwidth]{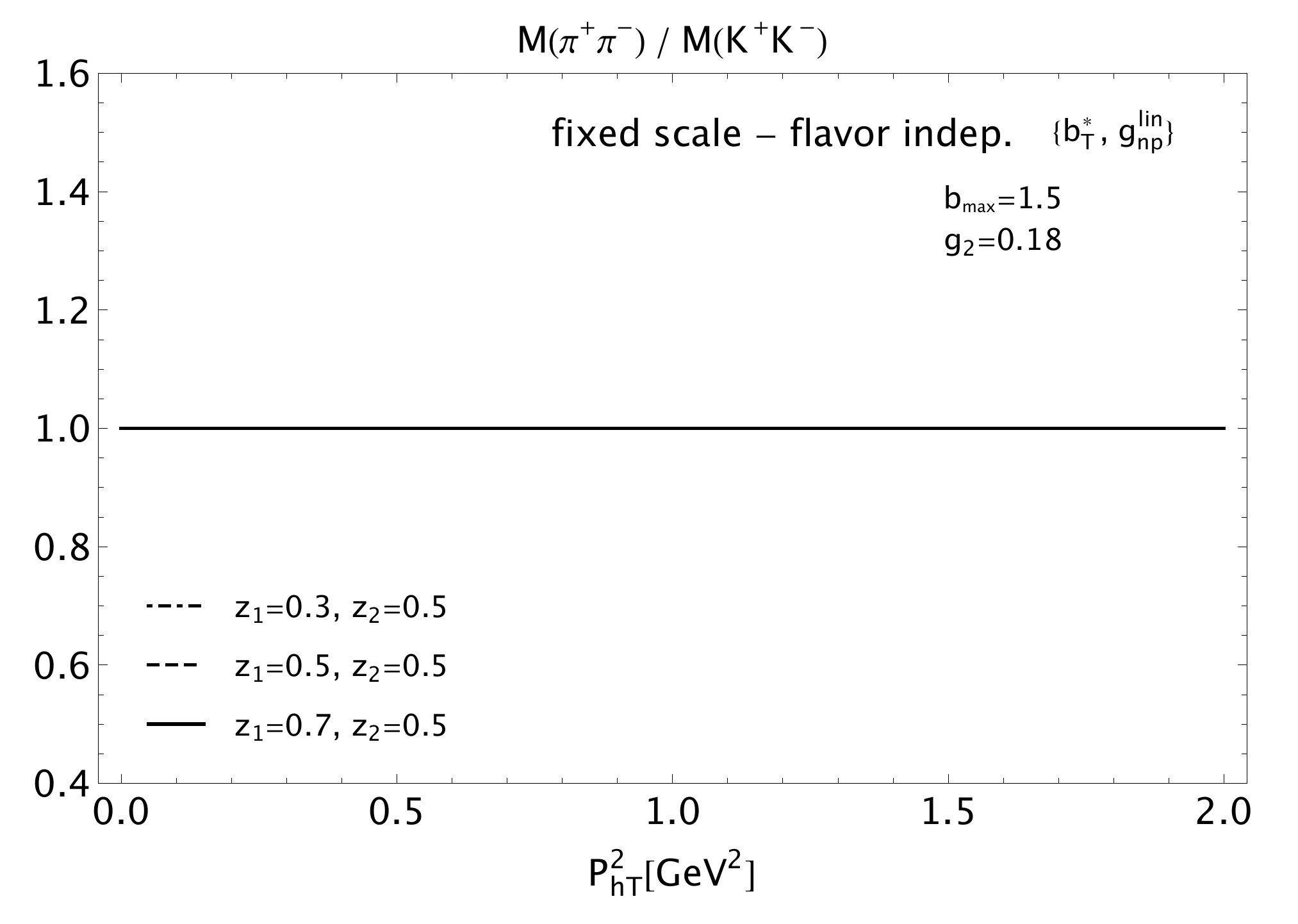}\hspace{.5cm} \includegraphics[width=0.4\textwidth]{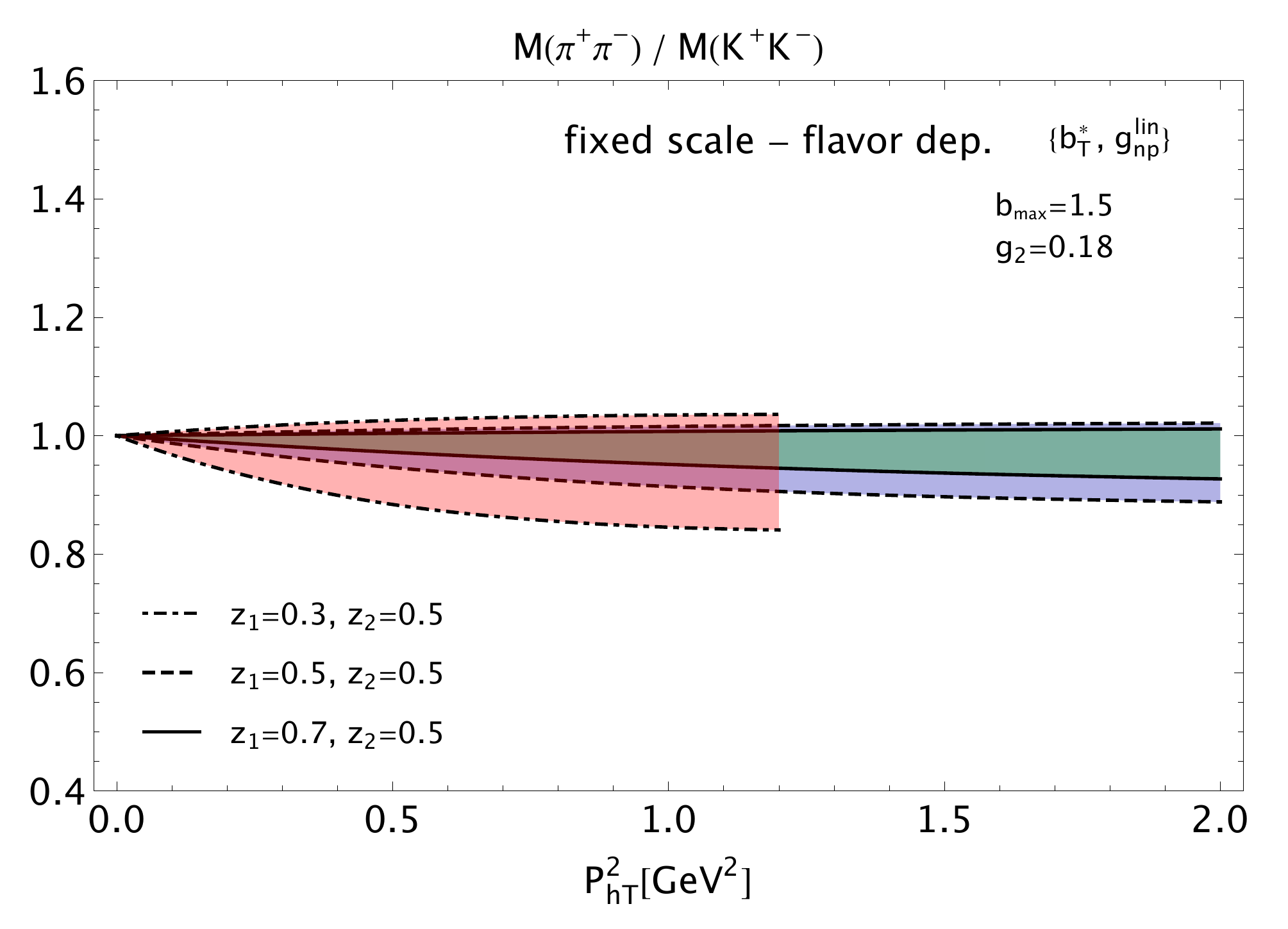}
  \caption{Tha ratio of normalized multiplicities in Eq.~(\ref{e:ratio_norm_mult}) between the $\{\pi^+ \pi^-\}$ final state and the $\{K^+ K^-\}$ final state at  $z_2=0.5$ and $y=0.2$ as a function of $\bm{P}_{1\p}^2 = z_1^2 \bm{q}_{\sT}^2$ at the \belle scale $Q^2=100$ GeV$^2$ for the "fixed scale" evolution scheme, for the evolution parameters $\{ \bmax = 1.5, \, g_2 = 0.18\}$, and with the $\{b_T^*, \, g_{\text{np}}^{\text{lin}}\}$ prescription for the transition to the nonperturbative regime (see text). Uncertainty bands with dot-dashed, dashed, and solid borders for $z_1=0.3, \, 0.5, \, 0.7,$ respectively. Left panel for  flavor independent intrinsic parameters of input TMD FF, right panel for flavor dependent ones (see text). }
  \label{f:ratio-fixed}
\eef

In Fig.~\ref{f:ratio-fixed}, we show the ratio of Eq.~(\ref{e:ratio_norm_mult}) between the normalized multiplicity for $\{ \pi^+ \pi^-\}$ and the one for $\{ K^+ K^-\}$ at $z_2=0.5$ and $y=0.2$ as a function of $\bm{P}_{1\p}^2 = z_1^2 \bm{q}_{\sT}^2$ at the \belle scale $Q^2=100$ GeV$^2$ for the "fixed scale" evolution scheme, for the evolution parameters $\{ \bmax = 1.5, \, g_2 = 0.18\}$, and with the $\{b_T^*, \, g_{\text{np}}^{\text{lin}}\}$ prescription for the transition to the nonperturbative regime. 

If we suppose to switch off the flavor dependence of the intrinsic parameters, the $b_{\sT}$ distribution of the TMD FF in Eq.~(\ref{e:hybrid}) is controlled by the same Gaussian width $\langle \bm{P}^2_{\p} \rangle (z)$ for all channels. This feature remains valid when performing the Bessel transform to momentum space, such that the $\bm{q}_{\sT}^2$ distribution of the cross section can be factorized out of the flavor sum. Therefore, if we take the ratio of normalized multiplicities at the same $z_1$ we expect the latter to be independent of $\bm{P}_{1\p}^2 = z_1^2 \bm{q}_{\sT}^2$. This is indeed the result displayed in the left panel of Fig.~\ref{f:ratio-fixed}. It is a systematic feature of the "fixed scale" evolution scheme: it holds true for other values of $z_1$, as shown in the panel, but also for other combinations of nonperturbative evolution parameters and nonperturbative prescriptions. 

If we account for the flavor dependence of the Gaussian widths $\langle \bm{P}^2_{\p} \rangle^{q\to h} (z)$, then the $b_{\sT}$ distribution is different for the $\{ \pi^+ \pi^-\}$ final state from the one for $\{ K^+ K^-\}$, as it can be realized by inspecting Eqs.~(\ref{e:favpi+pi-Gauss})-(\ref{e:unfK+K-Gauss}). Consequently, the ratio of normalized multiplicities has a specific $\bm{P}_{1\p}^2 = z_1^2 \bm{q}_{\sT}^2$ distribution that, of course, changes with $z_1$. This is indeed the content of the right panel in Fig.~\ref{f:ratio-fixed}: the uncertainty band of the 68\% of 200 replicas of Gaussian widths with dot-dashed borders corresponds to $z_1=0.3$, the band with dashed borders to $z_1=0.5$, the band with solid borders to $z_1=0.7$. 

Almost all the ratios are smaller than unity because in our  approximations the fragmentation into kaons has two favoured channels while the fragmentation into pions only one (see Eqs.~(\ref{e:favpi+pi-Gauss}) and (\ref{e:favK+K-Gauss})), and the $\bm{P}_{1\p}^2$ distribution of the fragmentation into kaons seems to be larger than the corresponding one for pions (see the analysis of Ref.~\cite{Signori:2013mda}). In any case, we believe that the inspection of the $\bm{P}_{1\p}^2$ distribution of ratios of normalized multiplicities for different final hadrons produced in future $e^+ e^-$ annihilation experiments is a useful tool to discriminate among different scenarios in TMD evolution. For example, if future data for this observable will lie well above unity, the "fixed scale" evolution scheme would be ruled out, independently of the flavor dependence of the intrinsic parameters in the TMD FF at the initial scale of evolution.

\bef
 \includegraphics[width=0.4\textwidth]{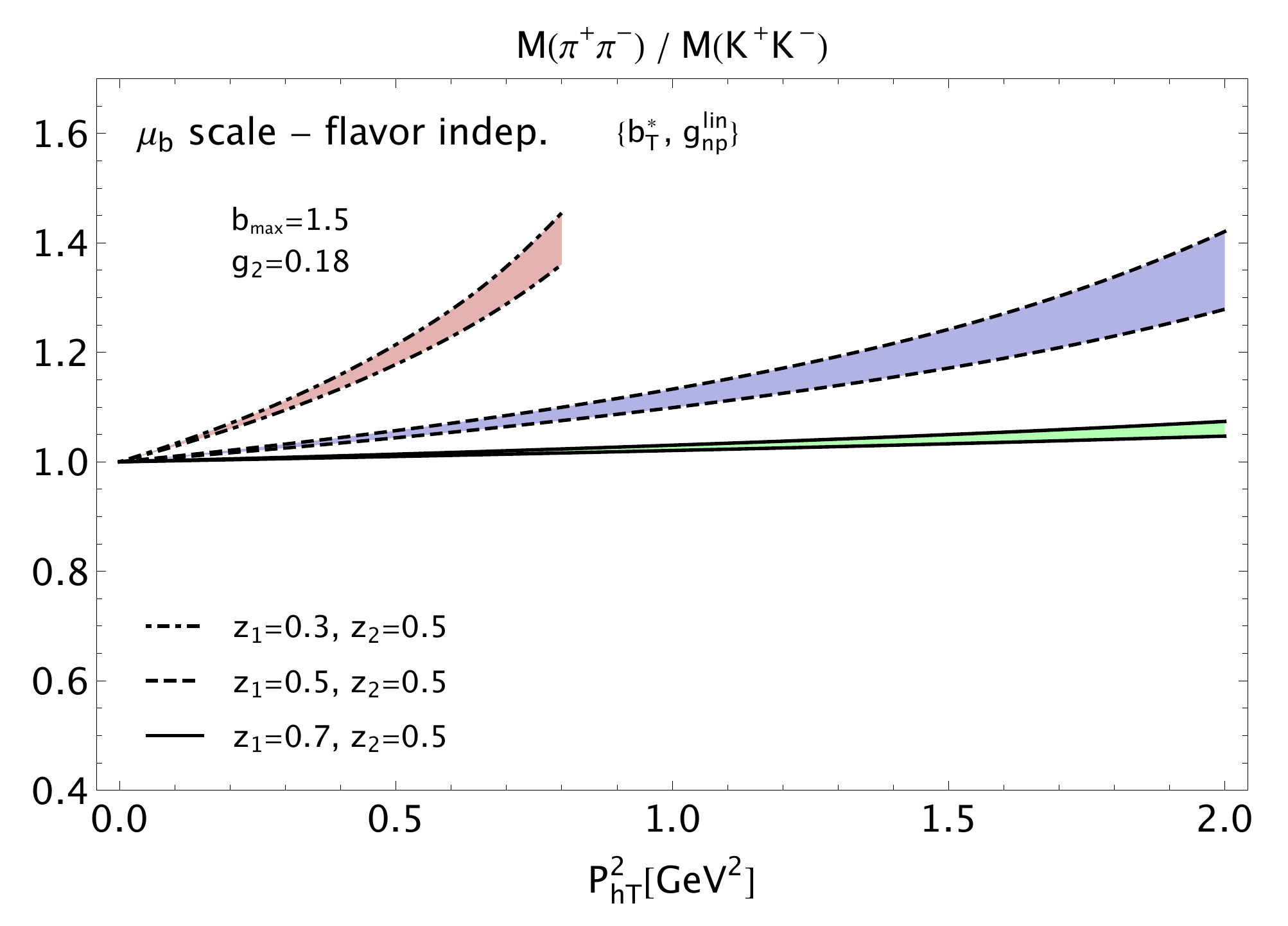}\hspace{.5cm} \includegraphics[width=0.4\textwidth]{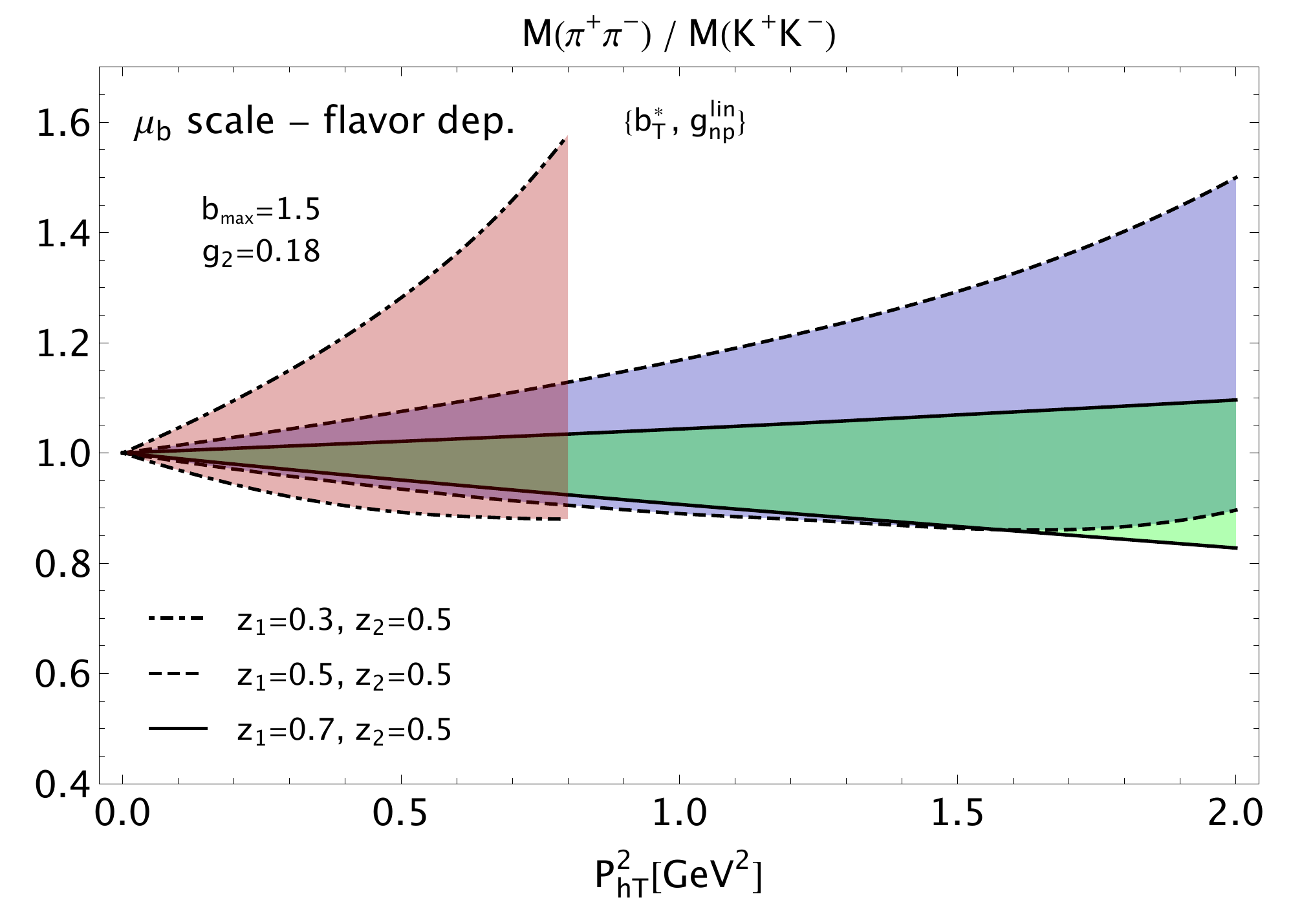} \\
 \centerline{\includegraphics[width=0.6\textwidth]{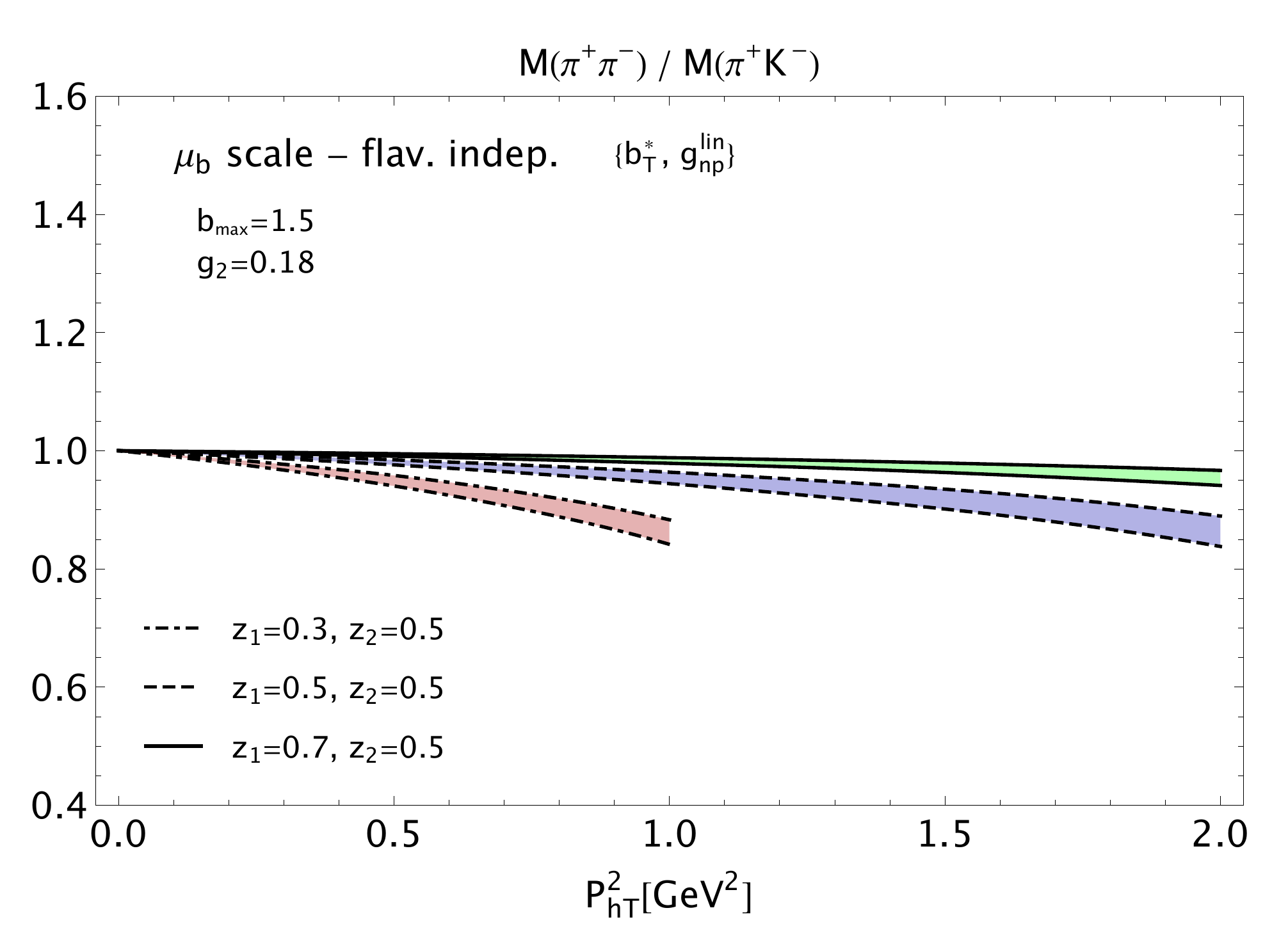}}
  \caption{Upper panels: same as in previous figure but for the "$\mu_b$ scale" evolution scheme. Lower panel: the ratio between the normalized multiplicities $M^{\pi^+ \pi^-} (z_1, z_2=0.5, q_{\sT}^2, y=0.2) / M^{\pi^+ \pi^-} (z_1, z_2=0.5, 0, y=0.2)$ and $M^{\pi^+ K^-} (z_1, z_2=0.5, q_{\sT}^2, y=0.2) / M^{\pi^+ K^-} (z_1, z_2=0.5, 0, y=0.2)$ as a function of $\bm{P}_{1\p}^2 = z_1^2 \bm{q}_{\sT}^2$ at the \belle scale $Q^2=100$ GeV$^2$ in the same conditions and with the same notation as in the upper panels, but for flavor independent intrinsic parameters of input TMD FF (see text). }
  \label{f:ratio-mub}
\eef

In Fig.~\ref{f:ratio-mub}, in the two panels of the upper row we show the same ratio of normalized multiplicities in the same conditions and notation as in the previous figure but for the "$\mb$ scale" evolution scheme. The left panel still corresponds to the case when the flavor dependence of the intrinsic parameters is neglected. However, in the "$\mb$ scale" scheme the $b_{\sT}$ distribution of the TMD FF is influenced also by the collinear part of the fragmentation function: the $d_1^{q\to h}$ in Eq.~(\ref{e:nCSS}) is evaluated at the running scale $\mu_{\hat{b}}$ which is related to $b_{\sT}$ via Eqs.~(\ref{e:def_mubhat}),~(\ref{e:bstar}),~(\ref{e:bsun}). Hence, when performing the Bessel transform of $D_1^q$ in the cross section, the resulting $\bm{q}_{\sT}^2$ distribution depends on the flavor of the fragmenting parton even if the intrinsic parameters do not. This "perturbative" flavor dependence, induced by RGE acting on the evolved collinear part of the TMD FF, mixes with the possible flavor dependence of the intrinsic parameters, making it rather difficult to disentangle the two effects. The left panel in the upper row shows the ratio of normalized multiplicities as a function of $\bm{P}_{1\p}^2 = z_1^2 \bm{q}_{\sT}^2$ for three different values of $z_1$. As in the previous figure, the band with dot-dashed borders corresponds to $z_1=0.3$, the band with dashed borders to $z_1=0.5$, and the band with solid borders to $z_1=0.7$. Suprisingly, all the ratios are larger than unity. When including also the flavor dependence in the intrinsic parameters, the uncertainty bands become larger because there is a marked sensitivity to all possible replica values of the intrinsic parameters themselves. Again, as in the previous section we can argue that experimental data will have a sufficiently small error to discriminate among the various replicas. 

A further constraint can be achieved by considering a different combination of final state hadrons in the ratio of normalized multiplicities in Eq.~(\ref{e:ratio_norm_mult}). The lower panel in Fig.~\ref{f:ratio-mub} shows the results for the ratio between a $\{ \pi^+ \pi^-\}$ final state and a $\{ \pi^+ K^-\}$ final state when neglecting the flavor dependence of intrinsic parameters of the TMD FF at the initial scale. The notation and conventions are the same as in the other panels. All the ratios are now lower than unity. Hence, combining this result with the content of the upper left panel could represent a very selective test of the "$\mb$ scale" evolution scheme. In fact, when neglecting the flavor dependence of intrinsic parameters the $\bm{P}_{1\p}^2$ distribution of normalized multiplicities for the $\{ \pi^+ \pi^-\}$ final state should be larger than the one for $\{ K^+ K^-\}$ at any $z_1$, while at the same time it should turn out narrower than the one for $\{ \pi^+ K^-\}$ at any $z_1$. Moreover, if future data for the $\{ \pi^+ \pi^-\}$ back-to-back production in $e^+ e^-$ annihilation will display a much narrower $\bm{P}_{1\p}^2$ distribution than for the $\{ K^+ K^-\}$ production, at least by 20\%, this will represent a further selective test for calculations performed in this evolution scheme, as it can be deduced by combining the results in the panels of the upper row.

Finally, we notice that because of charge conjugation symmetry (see Sec.~\ref{s:fav-unfav}) we predict that the ratio between normalized multiplicities leading to  $(\pi^+, \, K^-)$ and  $(\pi^-, \, K^+)$ final states should be equal to unity, irrespective of the choice of evolution schemes, nonperturbative evolution parameters and prescriptions. It would be interesting to cross-check this prediction by measuring this ratio as a function of $\bm{P}_{1\p}^2$. 

\section{Conclusions}
\label{s:end}

In this paper, we consider the semi-inclusive production of two back-to-back hadrons in electron-positron annihilations. We study the transverse momentum distribution of such pairs of hadrons by observing the mismatch between their collinear momenta, and we focus on charge-separated combinations of pions and kaons. We conveniently define the multiplicities in electron-positron annihilations as the differential number of back-to-back pairs of hadrons produced per corresponding single-hadron production, in analogy to the definition of multiplicity in SIDIS process. In particular, we analyze the multiplicities normalized to the point of vanishing transverse momentum in order to extract clean and uncontaminated details on the transverse momentum dependence of the functions describing the fragmentation process (transverse-momentum dependent fragmentation functions - TMD FFs). The normalized multiplicities are advantageous also because they turn out to be almost insensitive to the theoretical uncertainty related to the arbitrary choice of the renormalization scale. 

We consider electron-positron annihilations at large values of the center-of-mass (cm) energy, namely in the experimental conditions of the \belle and \bes experiments. We study how TMD FFs evolve with the hard scale. The input expression for TMD FFs is taken from a previous analysis of SIDIS multiplicities measured by \hermes at low energy, which is assumed as the starting scale. Since the hard scale in annihilation processes is much larger, we perform realistic tests on the sensitivity to various implementations of TMD evolution available in the literature. 

We find that within a specific evolution scheme the transverse momentum distribution of normalized multiplicities at the \belle scale can be very sensitive to the choice of the parameters describing the nonperturbative part of the evolution kernel. A hypothetical 7\% error in such data (compatible with the observed experimental error in collinear back-to-back emissions in electron-positron annihilations) could discriminate among different choices of parameters that are justified and adopted in the literature. 

But we observe also that at the same \belle scale different evolution schemes with different nonperturbative parameters can give overlapping transverse momentum distributions. Our global results indicate that different evolution schemes can be discriminated only by considering the combined dependence of normalized multiplicities on both the transverse momentum and the fractional energy carried by the final hadrons. And this finding holds true (with some limitations) also for the purpose of discriminating among different prescriptions for describing the transition from nonperturbative to perturbative regimes in transverse momentum. 

The dependence on the fractional energy of the final hadrons is contained in the collinear part of the TMD FFs. Different evolution schemes produce different evolution effects also in the collinear fragmentation functions, which in turn emphasize the differences in the final transverse momentum distribution of evolved TMD FFs. The dependence on the fractional energy is contained also in the average squared transverse momenta that describe the width of the input distribution of the TMD FFs at the starting scale. Therefore, by studying this dependence it may be possible to reduce the uncertainty on the intrinsic parameters that describe these input distributions. 

To this purpose, focusing on the normalized multiplicities at the \bes scale looks more promising. In fact, we observe that in stepping down from \belle to \bes scale the transverse momentum distributions of normalized multiplicities become much more sensitive to the details of the input distribution at the starting scale. The uncertainty in the determination of the intrinsic parameters needed to fit the \hermes SIDIS multiplicities reflects in a larger spread of normalized multiplicities as functions of transverse momentum. At the \bes scale, a hypothetical experimental error of 7\% does not discriminate among results coming from different nonperturbative evolution parameters or from different evolution schemes. But within a specific choice of evolution scheme it can discriminate among results that come from different values of the intrinsic parameters. 

The \hermes results al low energy show significant differences between SIDIS multiplicities for final-state pions and kaons. Hence, these data were fitted using transverse momentum distributions for the input TMD FFs that contain flavor dependent parameters. Here, we explore also how the final results for normalized multiplicities at \belle and \bes scales are sensitive to the details of this flavor dependence at the starting scale. In doing so, we find that the most convenient observable is represented by the ratio of normalized multiplicities for different final hadron species, particularly at the \belle scale. 

The most striking evidence is for evolution schemes where the flavor dependence is strictly localized only in the intrinsic parameters of the input TMD FFs at the starting scale. If we switch off such flavor dependence, the transverse momentum distribution of normalized multiplicities is always the same, irrespective of the species of final hadrons. So, if we select for example pions and kaons, the ratio of the corresponding normalized multiplicities is constant and equal to unity. If the flavor dependence of the intrinsic parameters is switched on, then the ratio deviates to values (mostly) lower than unity, in agreement with general expectations that kaons have a larger distribution in transverse momentum. 

The situation is more confused for evolution schemes where the flavor dependence is indirectly contained also in the initial conditions of the evolution equations through the (flavor dependent) collinear part of the fragmentation functions. In this case, this effect mixes up with the flavor dependence contained in the intrinsic transverse momentum distribution, and it is difficult to disentangle one from the other. At variance with the previous class of evolution schemes, in this case the ratio of normalized multiplicities for pions with respect to kaons turns out to be (mostly) larger than unity. Fortunately, more selective criteria are offered by considering a variety of species of final hadrons. If we consider ratios of normalized multiplicities for pions with respect to mixed pion-kaon pairs, the results are (mostly) lower than unity. By combining the results for various final states all together, one would hope to constrain the arbitrary ingredients of TMD FFs as much as possible. 

We conclude by stressing that all the results and remarks above refer to the unpolarized TMD FFs that describe the fragmentation of an unpolarized parton into an unpolarized hadron. However, this function is an essential ingredient in all the (spin) azimuthal asymmetries extracted in hard processes like electron-positron annihilation, hadronic collision, and SIDIS. Hence, a better control on the transverse momentum dependence of unpolarized TMD FFs implies also a better knowledge of polarized TMD FFs as well as of (un)polarized TMD parton distributions. For this reason, we are looking forward to a multidimensional analysis of data accumulated by the \belle and \bes collaborations, possibly including a study of normalized multiplicities for various hadron species as suggested in this work.

\section*{Acknowledgments}
Discussions with Christine Aidala, Ignazio Scimemi, Leonard Gamberg, Gunar Schnell, Charlotte van Hulse, Francesca Giordano, Isabella Garzia and Ted Rogers, are gratefully acknowledged.
The work of AS and MGE is part of the program of the Stichting voor Fundamenteel Onderzoek der Materie (FOM), which is financially supported by the Nederlandse Organisatie voor Wetenschappelijk Onderzoek (NWO).


\bibliographystyle{jhep}
\bibliography{biblio_epem}

\end{document}